\documentclass[prb,letterpaper,twocolumn,showpacs,superscriptaddress,floatfix,amsmath,amsfonts,amssymb,preprintnumbers,citeautoscript]{revtex4}
\pdfoutput=1 
\usepackage[T1]{fontenc}\usepackage[latin1]{inputenc}
\usepackage{dcolumn,graphicx,color,hvfloat,booktabs,microtype,afterpage} \graphicspath{{Figures/}}
\usepackage[charter]{mathdesign}
\renewcommand{\figurename}{Fig.}
\renewcommand{\tablename}{Table}
\makeatletter\renewcommand{\fnum@figure}[1]{\figurename~\thefigure~(color online).}\makeatother
\makeatletter\renewcommand{\fnum@table}[1]{\tablename~\thetable.}\makeatother
\usepackage[colorlinks,plainpages=false,linkcolor=blue,urlcolor=blue,citecolor=blue,pdfpagemode=UseNone,pdfstartview=FitBH]{hyperref}

\makeatletter
\newcommand{\bibstyle@tables}{\bibpunct[, ]{[}{]}{;}{n}{,}{, \hspace{-2.1pt}}%
    \gdef\bibnumfmt##1{[##1]}}
\makeatother

\begin{document}\pagestyle{plain}

\title{Crossover from weak to strong pairing in unconventional superconductors}

\author{D.\,S.\,Inosov}\email[\vspace{-0.9em}Corresponding author: \vspace{4pt}]{d.inosov@fkf.mpg.de}
\affiliation{\mbox{Max-Planck-Institut für Festkörperforschung, Heisenbergstraße 1, D-70569 Stuttgart, Germany}}

\author{J.\,T.~Park}
\affiliation{\mbox{Max-Planck-Institut für Festkörperforschung, Heisenbergstraße 1, D-70569 Stuttgart, Germany}}

\author{A.\,Charnukha}
\affiliation{\mbox{Max-Planck-Institut für Festkörperforschung, Heisenbergstraße 1, D-70569 Stuttgart, Germany}}

\author{Yuan~Li}
\affiliation{\mbox{Max-Planck-Institut für Festkörperforschung, Heisenbergstraße 1, D-70569 Stuttgart, Germany}}

\author{A.\,V.~Boris}
\affiliation{\mbox{Max-Planck-Institut für Festkörperforschung, Heisenbergstraße 1, D-70569 Stuttgart, Germany}}
\affiliation{\mbox{Department of Physics, Loughborough University, LE11\,3TU Loughborough, United Kingdom}}

\author{B.\,Keimer}
\affiliation{\mbox{Max-Planck-Institut für Festkörperforschung, Heisenbergstraße 1, D-70569 Stuttgart, Germany}}

\author{V.~Hinkov}
\affiliation{\mbox{Max-Planck-Institut für Festkörperforschung, Heisenbergstraße 1, D-70569 Stuttgart, Germany}}
\affiliation{Department of Physics and Astronomy, University of British Columbia, V6T\,1Z2 Vancouver, Canada}


\begin{abstract}
\noindent Superconductors are classified by their pairing mechanism and the coupling strength, measured as the ratio of the energy gap, $2\Delta$, to the critical temperature, $T_{\rm c}$. We present an extensive comparison of the $2\Delta/k_{\rm B}T_{\rm c}$ ratios among many single- and multiband superconductors from simple metals to high-$T_{\rm c}$ cuprates and iron pnictides. Contrary to the recently suggested universality of this ratio in Fe-based superconductors, we find that the coupling in pnictides ranges from weak, near the BCS limit, to strong, as in cuprates, bridging the gap between these two extremes. Moreover, for Fe- and Cu-based materials, our analysis reveals a universal correlation between the gap ratio and $T_{\rm c}$, which is not found in conventional superconductors and therefore supports a common unconventional pairing mechanism in both families. An important consequence of this result for ferropnictides is that the separation in energy between the excitonic spin-resonance mode and the particle-hole continuum, which determines the resonance damping, no longer appears independent of $T_{\rm c}$.
\end{abstract}

\maketitle\enlargethispage{3pt}

\vspace{-5pt}\section{Introduction}\vspace{-5pt}

\noindent At present, results of the few existing systematic experimental studies of the pairing strength in iron-arsenide superconductors remain at odds with each other. Some report a more or less universal value of $2\Delta/k_{\rm B}T_{\rm c}$, either below \cite{ZhangOh10} or well above \cite{EvtushinskyInosov09NJP, NakayamaSato11} the weak-coupling limit of 3.53 predicted by the Bardeen-Cooper-Schrieffer (BCS) theory, whereas others present evidence for a strongly doping-dependent coupling \cite{HardyBurger10}. The reported values of $2\Delta/k_{\rm B}T_{\rm c}$ scatter from as low as $\sim\kern1pt$3, below the weak-coupling limit \cite{HardyBurger10, KuritaRonning09, LiOoe10, ImaiTakahashi10, KimTanatar11}, to 10 and above \cite{TeagueDrayna11}, as summarized in Table~I in the Appendix. Hence, should one classify Fe-based superconductors as weakly or strongly coupled? Can they be at all considered as a single family?

To address these questions, we have analyzed all the available energy-gap reports in various Fe-based superconductors and their kin. We put these results into a broader context by comparing them to single- and multiband conventional superconductors, high-$T_{\rm c}$ cuprates, as well as heavy-fermion compounds and a few other superconducting (SC) materials. More than a hundred of such measurements are listed in Tables I\,--\,III (see Appendix).

\vspace{-5pt}\section{Gap ratios}\vspace{-5pt}

\begin{figure*}[t]\vspace{-1em}
\mbox{\includegraphics[width=\textwidth]{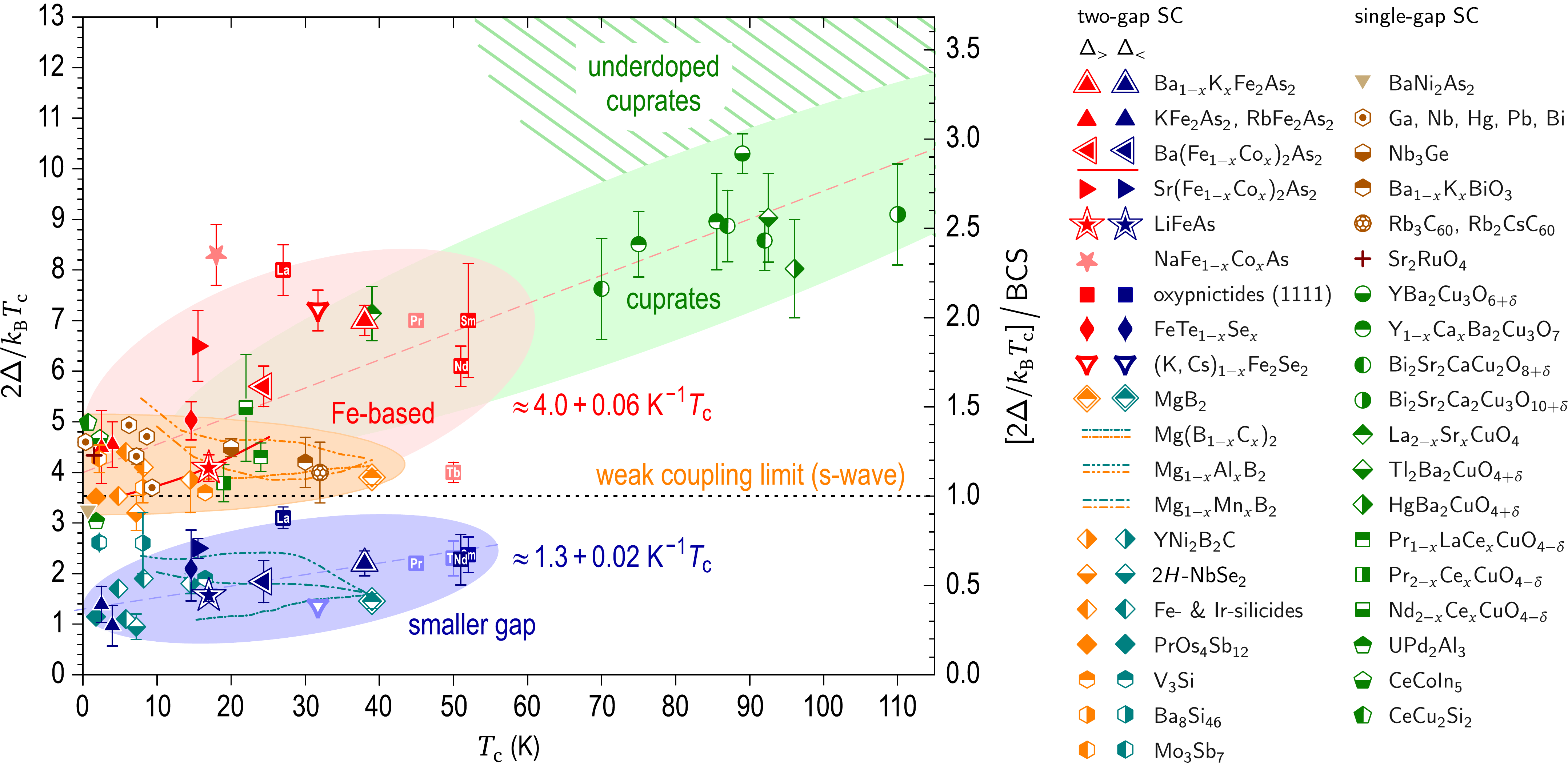}}
\caption{The gap ratios, $2\Delta/k_{\rm B}T_{\rm c}$, for different families of single- and two-gap superconductors vs. their critical temperatures at ambient pressure, $T_{\rm c}$. The data points summarize most of the recent energy-gap measurements in ferropnictides, high-$T_{\rm c}$ cuprates \cite{YuLi09}, and some conventional superconductors. Each data point is an average of all the available measurements of the corresponding compound by various complementary techniques (see Tables I\,--\,III). The error bars represent one standard deviation of this average for repeatedly measured compounds or the experimental errors of single measurements, whenever averaging could not be performed. Such unconfirmed points are shown in lighter colors. Points confirmed in a considerable number of complementary measurements are additionally outlined. The weak-coupling limit, predicted for s-wave superconductors by the BCS theory, is shown by the dotted line. For weakly coupled d-wave superconductors, a slightly higher value of 4.12 is expected (not shown).\vspace{-1.2em}}
\label{Fig:GapTc}
\end{figure*}

Fe-based superconductors are multiband metals, whose conduction bands are formed almost exclusively by the Fe\,3d electrons \cite{KoitzschInosov08, AndersenBoeri11}. Because in the SC state they typically exhibit energy gaps of two sizes \cite{EvtushinskyInosov09NJP, CharnukhaDolgov11, ChubukovEfremov08, MatanoRen08, PopovichBoris10, DingRichard08, EvtushinskyInosov09}, it is illustrative to compare them to other multigap superconductors, such as MgB$_2$ \cite{SzaboSamuely01, GonnelliDaghero02, GiubileoRoditchev02, IavaroneKarapetrov05, ChenKonstantinovic01, TsudaYokoya01, TsudaYokoya03, SoumaMachida03, Xi08}, as well as to the high-$T_{\rm c}$ materials with a single gap. In Fig.\,\ref{Fig:GapTc}, the gap ratios, $2\Delta/k_{\rm B}T_{\rm c}$, are plotted vs. $T_{\rm c}$. For multigap superconductors, we differentiate between the small ($\Delta_<$) and large ($\Delta_>$) energy gaps, which lie below and above the weak-coupling limit, respectively \cite{DolgovMazin09}.

First of all, we note that the majority of low-$T_{\rm c}$ superconductors, including heavy-fermion compounds, such as CeCoIn$_5$, CeCu$_2$Si$_2$ or UPd$_2$Al$_3$, exhibit relatively low gap ratios within $\sim$\,30\,\% of the BCS limit, according to the latest reports \cite{ParkGreene09, FujiwaraHata08}. In conventional superconductors, the gap ratios remain in this narrow range (semielliptical shaded region in Fig.\,\ref{Fig:GapTc}) even at higher $T_{\rm c}$, as best illustrated by Ba$_{1-x}$K$_x$BiO$_3$ ($T_{\rm c}=30$\,K) \cite{SatoTakagi90, PuchkovTimusk94, MarsiglioCarbotte96}, Rb$_2$CsC$_{60}$ ($T_{\rm c}=33$\,K) \cite{ForroMihaly01} or MgB$_2$ ($T_{\rm c}=39$\,K) with its chemically substituted derivatives \cite{GonnelliDaghero02, KortusDolgov05, GonnelliDaghero07, Xi08}. This behavior is in stark contrast to that of unconventional superconductors, such as Fe-based compounds or over- and optimally-doped copper oxides. There, the $2\Delta_>/k_{\rm B}T_{\rm c}$ ratios exhibit a statistically significant positive correlation with $T_{\rm c}$ and for the majority of materials cluster along the $4.0+0.06\,{\rm K}^{-1}T_{\rm c}$ line, shared by both families. This universal behavior could result from a common pairing mechanism in these two families that clearly differentiates them from phonon-mediated superconductors. Underdoped cuprates, however, do not conform to this scaling and exhibit even higher $2\Delta/k_{\rm B}T_{\rm c}$ ratios (hatched region in Fig.\,\ref{Fig:GapTc}) due to the influence of the pseudogap and proximity to the Mott-insulating state. Therefore, we have restricted our collection of cuprates to over- and optimally-doped compounds, where superconductivity is not impaired by any competing phases.

\makeatletter\renewcommand{\fnum@figure}[1]{\figurename~\thefigure~(color online).}\makeatother

A closer look at the Fe-based superconductors reveals a wide spread of gap ratios, from weak BCS-like values in non-magnetic LiFeAs \cite{BorisenkoZabolotnyy10, InosovWhite10, SasmalLv10, SongGhim11, WeiChen10, StockertAbdelHafiez10} to twice larger values in high-$T_{\rm c}$ ferropnictides with strong antiferromagnetic (AFM) correlations, such as optimally-doped Ba$_{1-x}$K$_x$Fe$_2$As$_2$ (BKFA) \cite{PopovichBoris10, DingRichard08, EvtushinskyInosov09, NakayamaSato09, XuHuang11, LuPark10, SzaboPribulova09, HiraishiKadono09, KhasanovEvtushinsky09} or various 1111-compounds \cite{KawasakiShimada08, GonnelliDaghero09, DagheroTortello09, MatanoRen08, KondoSantanderSyro08, SamuelySzabo09, GonnelliDaghero09a}. We would like to emphasize that despite all the uncertainties in the published values, these differences are established beyond any doubt, as they have been confirmed by many complementary experiments, at least for several most studied materials (Table~I). Therefore, in contrast to the high-$T_{\rm c}$ cuprates, which can be generally classified as strong-coupling superconductors, Fe-based systems show a larger variability and fill in the wide gap between conventional and cupratelike pairing strengths. The overall trend confirms that the superlinear increase of $\Delta_>$ with $T_{\rm c}$, suggested in Ref.~\citenum{HardyBurger10}, remains qualitatively valid for all Fe-based compounds in general. However, the absolute values of the gap ratios for Ba(Fe$_{1-x}$Co$_x$)$_2$As$_2$ (BFCA), extracted from heat capacity measurements in Ref.~\citenum{HardyBurger10} (solid line in Fig.\,\ref{Fig:GapTc}), appear to be somewhat underestimated in comparison to other reports.

Next, we consider the smaller gap, which is found in many multiband superconductors below the BCS limit. For all studied superconductors (both conventional and unconventional), we find somewhat smaller variability of the $2\Delta_</k_{\rm B}T_{\rm c}$ values, which tend to accumulate close to the $1.3+0.02\,{\rm K}^{-1}T_{\rm c}$ line. The fact that its slope has the same sign as that for the larger gap is consistent with predictions of the Eliashberg theory for interband pairing \cite{DolgovMazin09, CharnukhaDolgov11}, suggesting a similar scaling of both gaps with the effective coupling ($\lambda_{\rm eff}$ in Ref.~\citenum{DolgovMazin09}), in contrast to the BCS formalism.
\enlargethispage{2pt}

Let us now discuss several particular test cases for the above-mentioned trends. The first example comes from the juxtaposition of the stoichiometric conventional superconductor MgB$_2$ ($T_{\rm c}=39$\,K) \cite{SzaboSamuely01, GonnelliDaghero02, GiubileoRoditchev02, IavaroneKarapetrov05, ChenKonstantinovic01, TsudaYokoya01, TsudaYokoya03, SoumaMachida03, Xi08} and the optimally hole-doped BKFA ($T_{\rm c,\,max}=38.5$\,K) \cite{PopovichBoris10, DingRichard08, EvtushinskyInosov09, NakayamaSato09, XuHuang11, LuPark10, SzaboPribulova09, HiraishiKadono09, KhasanovEvtushinsky09, CharnukhaPopovich11}. Both are multiband superconductors with almost identical critical temperatures, and their two well-separated SC gaps have been extensively measured by various experimental methods, such as angle-resolved photoemission (ARPES) \cite{DingRichard08, EvtushinskyInosov09, NakayamaSato09, XuHuang11, ZhaoLiu08}, scanning tunneling spectroscopy (STS) \cite{ShanWang11, ShanWang11nphys, WrayQian08}, point-contact Andreev reflection (PCAR) spectroscopy \cite{LuPark10, SzaboPribulova09}, muon-spin rotation ($\mu$SR) \cite{HiraishiKadono09, KhasanovEvtushinsky09}, calorimetry \cite{PopovichBoris10}, and others (see Table~I). By averaging these results, the gap ratios can be determined with a very small uncertainty. The larger gap in MgB$_2$ yields an average $2\Delta_>/k_{\rm B}T_{\rm c}$ ratio of 3.9\,$\pm$\,0.13, only 10\,\% above the weak-coupling limit \cite{GonnelliDaghero02, KortusDolgov05, GonnelliDaghero07, Xi08}. The corresponding ratio for BKFA, however, is 7.0\,$\pm$\,0.3, almost twice the BCS value. For the smaller gap, we find a qualitatively similar difference.

It is tempting to ascribe this difference to the stronger coupling in ferropnictides in general, but such a scenario is disproved by our second test case, where we compare differently doped Ba-122 materials. Superconductivity in the Ba-122 family can be induced either by a partial substitution of Ba with K or Rb that leads to hole doping of the FeAs layers, or by replacing Fe atoms with Co or Ni within the layers. The end points of both series, corresponding to 100\,\% substitution, are stoichiometric low-$T_{\rm c}$ superconductors KFe$_2$As$_2$ ($T_{\rm c}=4$\,K), RbFe$_2$As$_2$ ($T_{\rm c}=2.5$\,K) and BaNi$_2$As$_2$ ($T_{\rm c}=0.68$\,K), all characterized by weak coupling \cite{KuritaRonning09, ShermadiniKanter10, KawanoFurukawa10}. Moreover, BaNi$_2$As$_2$ appears to be a conventional phonon-mediated superconductor \cite{KuritaRonning09, RonningBauer09}. This implies that the $2\Delta_>/k_{\rm B}T_{\rm c}$ ratio must vary continuously with doping within the Ba-122 family\,---\,an effect that so far has been directly observed only in the Co-doped series \cite{HardyBurger10}. Fig.\,\ref{Fig:GapTc} suggests this variation to be even stronger (almost twofold) in BKFA, where higher values of $T_{\rm c}$ can be reached. Indeed, the extensively studied optimally-doped BFCA ($T_{\rm c}=25$\,K) has an average gap ratio of only 5.4\,$\pm$\,0.4, in the middle between those of optimally-doped BKFA and weakly coupled superconductors \cite{SamuelyPribulova09, TerashimaSekiba09, YinZech09}.

To complete our chain of comparisons, we now focus on the high-$T_{\rm c}$ part of the plot that contains oxypnictides and most of the copper oxides. With the exception of a single, so far unconfirmed, PCAR measurement on Tb-1111 \cite{YatesMorrison09}, most other works report high values of the gap ratios in La-, Pr-, Nd-, and Sm-based 1111 compounds \cite{KawasakiShimada08, GonnelliDaghero09, DagheroTortello09, MatanoRen08, KondoSantanderSyro08, SamuelySzabo09, GonnelliDaghero09a}, with an average around 7\,$\pm$\,1. In high-$T_{\rm c}$ copper oxides with similar or slightly higher critical temperatures, such as Bi$_2$Sr$_2$CaCu$_2$O$_{8+\delta}$ (Bi-2212), comparable ratios around 8.5\,$\pm$\,0.5 have been reported \cite{FedorovValla99, BorisenkoKordyuk02, LeeVishik07} (see Table~III). A further increase of the $2\Delta/k_{\rm B}T_{\rm c}$ ratio towards $\sim$\,10, close to the strong-coupling limit of the Eliashberg theory \cite{CombescotDolgov96}, is observed in Hg-1223 ($T_{\rm c}$\,=\,130\,K) and Hg-1201 ($T_{\rm c}$\,=\,96\,K) cuprates \cite{YuLi09}, suggesting that the positive correlation between this ratio and $T_{\rm c}$, similar to the one we found for Fe-based compounds, could be universal for all unconventional superconductors, including cuprates. Gap ratios in the most recently discovered iron-selenide superconductors \cite{GuoJin10, MizuguchiTakeya11, KrztonMaziopa11} ($T_{\rm c,\,max}\approx33$\,K) also conform to this general trend \cite{YuanDong11, WangQian11, ZhaoMou11, MouLiu11, ZhangYang11, YuMa11} and are similar to those of optimally-doped BKFA.

However, we cannot fail to mention some deviations from this trend that are best demonstrated by LiFeAs, the bearer of the highest known $T_{\rm c}=18$\,K among stoichiometric Fe-based materials, together with its close relative NaFeAs. Despite its relatively high $T_{\rm c}$, LiFeAs is characterized by weak coupling barely above the BCS limit \cite{BorisenkoZabolotnyy10, InosovWhite10, SasmalLv10, SongGhim11, WeiChen10, StockertAbdelHafiez10, Hanaguri11, LiKawasaki11}, possibly related to the absence of notable Fermi surface nesting in its band structure \cite{BorisenkoZabolotnyy10} or even a different pairing mechanism \cite{KordyukZabolotnyy11, BrydonDaghofer11}. In NaFeAs, on the contrary, superconductivity with $T_{\rm c} \approx 10$\,K coexists with antiferromagnetism \cite{ParkerSmith10}. Upon electron doping, the AFM order is destroyed and critical temperatures up to 20\,K can be reached, resulting in a phase diagram \cite{ParkerSmith10} similar to those of 122-ferropnictides, in which the SC dome envelops an AFM quantum critical point. The SC gap in slightly overdoped NaFe$_{0.95}$Co$_{0.05}$As ($T_{\rm c} = 18$\,K, i.e.~coinciding with that of LiFeAs) was recently measured by ARPES \cite{LiuRichard10}, resulting in $2\Delta/k_{\rm B}T_{\rm c}=8.3\pm0.6$, which is much higher than in LiFeAs. This example illustrates that despite the above-mentioned correlation between the gap ratio and $T_{\rm c}$, identical critical temperatures even among Fe-based superconductors can still correspond to $2\Delta/k_{\rm B}T_{\rm c}$ values as different as those of MgB$_2$ and optimally-doped BKFA that we compared earlier. The relative role of magnetic correlations, doping-induced inhomogeneities, exotic pairing mechanisms, and other factors possibly leading to this exceptional behavior still remains to be investigated.

\makeatletter\renewcommand{\fnum@figure}[1]{\figurename~\thefigure.}\makeatother
\begin{figure}[t]
\mbox{\includegraphics[width=0.5\textwidth]{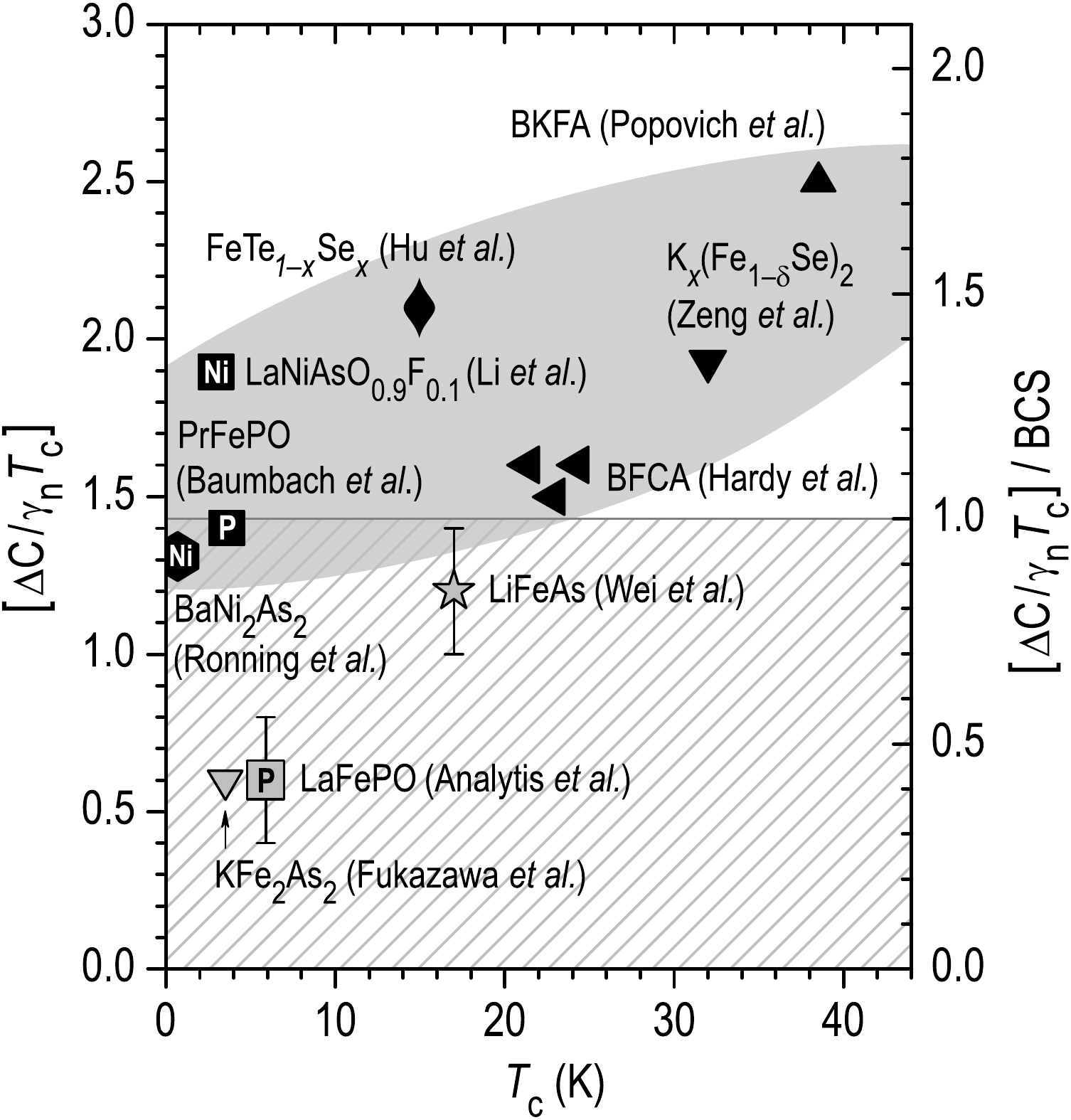}}
\caption{Heat-capacity measurements of the ${\scriptstyle\Delta}C/\gamma_{\rm n} T_{\rm c}$ ratio for various Fe-based superconductors (see Table~IV), as compared to the BCS prediction of 1.43. Points that were reported below this weak-coupling limit (hatched area) are shown in grey color.}
\label{Fig:SpecificHeat}\vspace{-2em}
\end{figure}

\vspace{-5pt}\section{Heat-capacity jump}\vspace{-5pt}

Energy-gap measurements are not the only way to quantify the deviation of a superconductor from the weak-coupling limit. Calorimetry provides direct access to the magnitude of the jump, ${\scriptstyle\Delta}C$, in the electronic specific heat at $T_{\rm c}$ (for a review in iron pnictides, see Ref.~\citenum{PaglioneGreene10}). In the framework of the BCS theory, it is related to the normal-state Sommerfeld coefficient, $\gamma_{\rm n}$, by ${\scriptstyle\Delta}C/\gamma_{\rm n} T_{\rm c}=1.43$, whereas in conventional superconductors with stronger coupling this ratio was shown to increase monotonically with $2\Delta/k_{\rm B}T_{\rm c}$~\cite{Carbotte90}. In Fig.\,\ref{Fig:SpecificHeat}, we compare the specific-heat-jump ratio reported in some Fe-based superconductors \cite{PopovichBoris10, FukazawaYamada09, HardyWolf10, HardyBurger10, RonningKurita08, WeiChen10, StockertAbdelHafiez10, BaumbachHamlin09, AnalytisChu08, LiChen08, ZengShen11}. For optimally doped BKFA with a relatively high value of $T_{\rm c}$, the ${\scriptstyle\Delta}C/\gamma_{\rm n} T_{\rm c}$ ratio lies 75\,\% above the BCS limit \cite{PopovichBoris10}. It exceeds all other values reported for pnictides with lower critical temperatures, confirming the increased deviation from the BCS prediction as $T_{\rm c}$ increases.

\vspace{-5pt}\section{Spin-resonance mode: scaling relationships}\vspace{-5pt}

It is remarkable that the largest deviations from the BCS limit are found in those compounds that possess an intense spectrum of spin fluctuations, which are believed to be important for the SC pairing. In contrast to the phonon spectrum, which is to a good approximation insensitive to the SC transition, magnetic excitations originate within the electronic subsystem and may experience drastic changes below $T_{\rm c}$, manifest in the spectral weight redistribution and the formation of a spin-resonance mode both in high-$T_{\rm c}$ cuprates \cite{RossatMignod91, FongKeimer95, FongBourges99, YuLi09} and in ferropnictides \cite{InosovPark10, WangLuo10, ParkInosov10, ShamotoIshikado10, ZhangWang11, ChristiansonGoremychkin08, ChristiansonLumsden09, PrattKreyssig10, LumsdenChristianson09, ChiSchneidewind09, LiChen09, ZhaoRegnault10, IshikadoNagai11, QiuBao09, ArgyriouHiess10, WenXu10, MookLumsden10, MookLumsden10a}. Such changes could offer a positive feedback effect, stabilizing the SC state and contributing to the excessively large gap amplitudes.

\makeatletter\renewcommand{\fnum@figure}[1]{\figurename~\thefigure~(color online).}\makeatother
\begin{figure}[t]
\mbox{\includegraphics[width=\columnwidth]{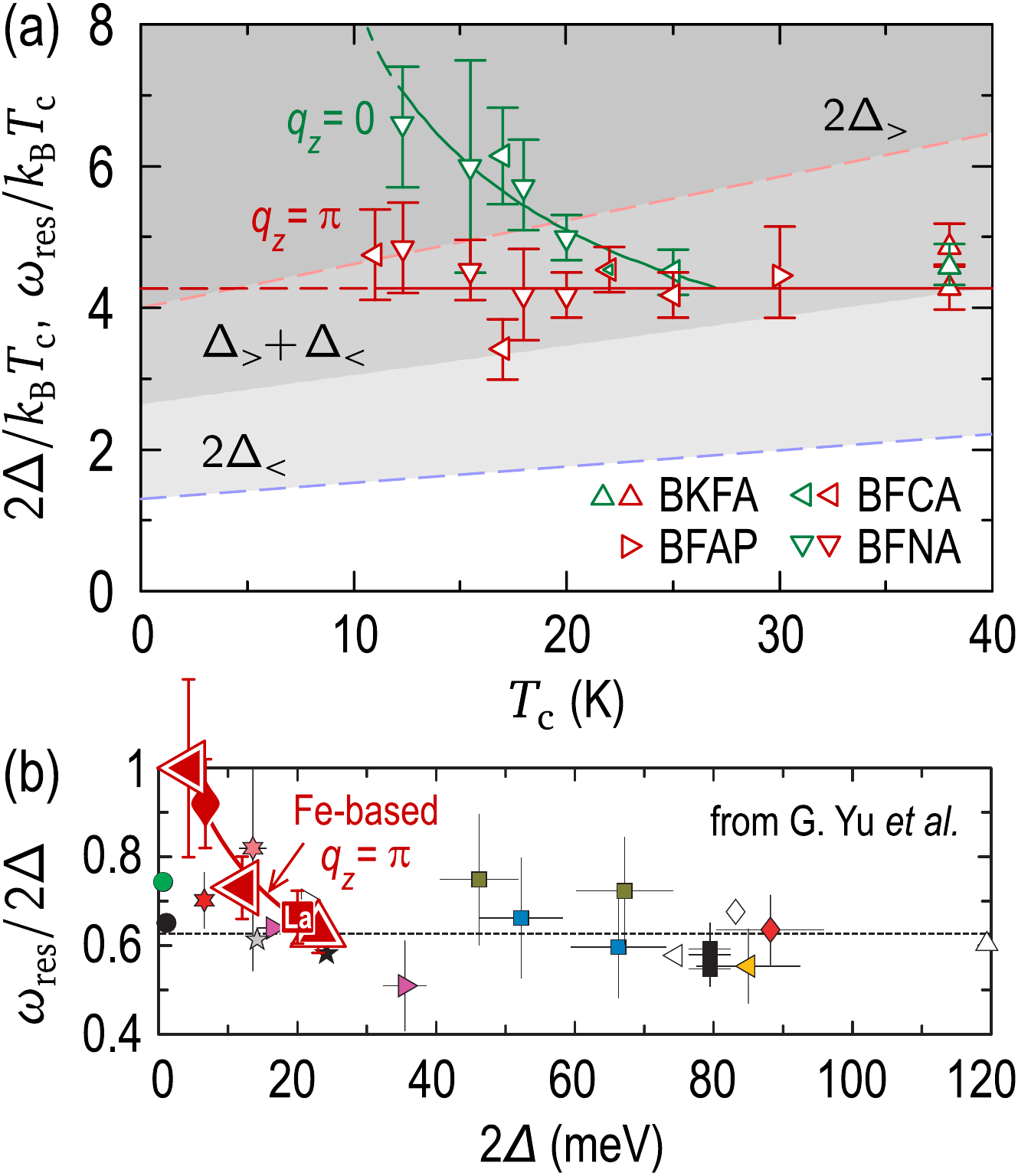}}
\caption{(a)~Normalized spin-resonance energy, $\omega_{\rm res}/k_{\rm B}T_{\rm c}$, in the Ba-122 iron arsenides for $q_z=\pi$ and $q_z=0$ (Ref.\,\citenum{InosovPark10, WangLuo10, ParkInosov10, ShamotoIshikado10, ZhangWang11}), plotted vs.\,$T_{\rm c}$ (see Table~V). The gray shading shows the particle-hole continuum with a three-step onset at $2\Delta_<$, $\Delta_<\!+\Delta_>$ and $2\Delta_>$. (b)~Ratios of the spin-resonance energy at $q_{\rm z}\!=\!\piup$ to the SC gap, $\omega_{\rm res}/2\Delta$, in Fe-based superconductors (large symbols) in comparison to the universal ratio of 0.64 proposed for other unconventional superconductors \cite{YuLi09}. The meaning of the symbols is retained from Fig.~\ref{Fig:GapTc} and Ref.~\citenum{YuLi09}, respectively.\vspace{-1.3em}}
\label{Fig:GapRes}
\end{figure}

Conversely, the proximity of the spin-excitonic resonance to $2\Delta$ determines its damping by particle-hole scattering \cite{PailhesUlrich06}, hence the behavior of the energy gap discussed above has important consequences for the SC resonant mode. In 122-compounds, its energy, $\omega_{\rm res}$, varies with the out-of-plane component of the momentum, $q_z$, so that its minimum, reached at $q_z=\piup$, scales linearly with $T_{\rm c}$, whereas the maximal value at $q_z=0$ always stays above 4\,meV, if extrapolated down to $T_{\rm c}\rightarrow0$ \cite{InosovPark10, WangLuo10, ParkInosov10, ShamotoIshikado10, ZhangWang11}. This results in $\omega_{\rm res}/k_{\rm B}T_{\rm c}$ ratios that are plotted in Fig.\,\ref{Fig:GapRes}\,(a). The ratio stays constant for $q_z=\piup$, but diverges for $q_z=0$ as $T_{\rm c}\rightarrow0$. Because $2\Delta/k_{\rm B}T_{\rm c}$ remains finite at all temperatures, such a behavior must increasingly suppress the resonance intensity for $q_z=0$ as its energy enters the particle-hole continuum [shaded regions in Fig.\,\ref{Fig:GapRes}\,(a)] with decreasing $T_{\rm c}$. So far, direct experimental evidence for such a suppression \cite{PrattKreyssig10} remain scarce. A systematic investigation of the resonant peak's intensity and shape for doping levels with $T_{\rm c}<11$\,K is therefore warranted.

For $q_z\!=\!\piup$, the situation with the resonance damping is more speculative, as it depends on the detailed $q_z$-dispersion of the continuum and the exact $T_{\rm c}$-dependence of the gap ratios. Generally for a two-gap superconductor, the particle-hole continuum has a three-step onset at $2\Delta_<$, $\Delta_<\!+\Delta_>$ and $2\Delta_>$. In 122-superconductors, however, the smaller gap typically resides only on one of the $\Gamma$-centered holelike bands \cite{DingRichard08, EvtushinskyInosov09, NakayamaSato09, XuHuang11}, rendering $2\Delta_<$ onset irrelevant for interband scattering close to the nesting vector. In electron-doped 122-compounds with optimal $T_{\rm c}$, the resonance mode appears below $2\Delta_>$, but has a significant overlap with $\Delta_<\!+\Delta_>$, which possibly contributes to its unusually large energy width \cite{InosovPark10, WangLuo10, ParkInosov10, ShamotoIshikado10, ZhangWang11}. This situation would not change with doping under the assumption of constant $2\Delta_>/k_{\rm B}T_{\rm c}$ ratios. However, if one assumes them to follow the average linear trends implied by Fig.\,\ref{Fig:GapTc} (dashed lines), the resonance would approach $2\Delta_>$ even at $q_z=\piup$, leading to its further broadening and suppression. This possibility is consistent with the fact that resonant modes have not so far been reported in either under- or overdoped samples with $T_{\rm c} < 11$\,K.

The described behavior of the gap implies that Fe-based superconductors violate the universality of the $\omega_{\rm res}/2\Delta$ ratio proposed in Ref.~\citenum{YuLi09}. Indeed, according to gap values in Fig.\,\ref{Fig:GapTc} and the proportionality $\omega_{\rm res}\approx(4.6\pm0.4)\,k_{\rm B}T_{\rm c}$, established in Ref.~\citenum{InosovPark10, WangLuo10, ParkInosov10, ShamotoIshikado10, ZhangWang11}, this ratio continuously increases from $\sim$\,0.65 in the optimally doped BKFA to $\sim$\,0.8 in the optimally doped BFCA. Then it approaches unity in compounds with even lower $T_{\rm c}$, such as underdoped BFCA or the 11-family, as illustrated by the large red symbols in Fig.\,\ref{Fig:GapRes}\,(b). The universal ratio of $\omega_{\rm res}/2\Delta=0.64$ has been interpreted as the result of a fundamental spin-mediated pairing mechanism in unconventional superconductors \cite{YuLi09}. Therefore, its breakdown in Fe-based systems, which becomes increasingly pronounced for low-$T_{\rm c}$ compounds (Table~V), might be indicative of a variation in the role played by spin fluctuations. Supposedly, they become increasingly less important to the SC pairing as $T_{\rm c}$ decreases (e.g. due to an interplay with conventional phononic pairing), which can explain the simultaneous increase in $\omega_{\rm res}/2\Delta$ and the reduction of the gap ratio.

Recently we became aware of a new inelastic-neutron-scattering (INS) study \cite{CastellanRosenkranz11} performed on several overdoped samples of polycrystalline BKFA. The results of this work indicate that the deviation of the $\omega_{\rm res}/2\Delta$ ratio from the ``universal'' value \cite{YuLi09} and the suppression of the resonant-mode spectral weight with decreasing $T_{\rm c}$, discussed above, also hold on the overdoped side of the phase diagram.

Another recent work \cite{TaylorPitcher11} has lately revealed an enhancement of the antiferromagnetic INS signal in LiFeAs below $T_{\rm c}$, resembling an overdamped spin-resonance mode. It is strongly broadened in energy and appears centered around $\sim$\,8\,meV, i.e. above $2\Delta\approx6.1\pm0.5$\,meV (see Table~V). This implies a considerable overlap of the resonance peak with the particle-hole continuum (as in under- or overdoped 122-systems) and a large $\omega_{\rm res}/2\Delta$ ratio of $1.3\pm0.4$, far above the ``universal'' value of 0.64. The results are consistent with the weak-coupling behavior suggested earlier by the small gap ratios observed in this compound \cite{BorisenkoZabolotnyy10, InosovWhite10, SasmalLv10, SongGhim11, WeiChen10, StockertAbdelHafiez10}.

\vspace{-5pt}\section*{Acknowledgments}\vspace{-5pt}

This work has been supported, in part, by the DFG within the Schwerpunktprogramm 1458, under Grant No.~\mbox{BO3537/1-1}, and by the MPI\,--\,UBC Center for Quantum Materials. The authors are grateful to L.\,Boeri, O.\,V.~Dolgov, D.\,V.~Efremov, D.\,V.~Evtushinsky, and R.\,Osborn for stimulating discussions.
\clearpage

\citestyle{tables}

\makeatletter\immediate\write\@auxout{\string\bibstyle{my-apsrev}}\makeatother

\onecolumngrid

\vspace*{-6em}\section*{APPENDIX: TABLES}

{\flushleft\footnotesize
\begin{tabular}[c]{@{}l@{~}l@{~~}l@{~~}l@{~~}l@{\hspace{1pt}}l@{~~}l@{\hspace{1pt}}l@{\quad}l@{\hspace{1pt}}l@{~~}l@{\hspace{1pt}}l@{~~}l@{~~}l@{\hspace{-3.2em}}r@{}}
\toprule
\multicolumn{2}{l}{\!Doping level} & sample & $T_{\rm c}$\,(K) & \multicolumn{2}{l}{$\kern-2pt\Delta_<$\,(meV)} & \multicolumn{2}{l}{\!\!$2\Delta_</k_{\rm B}T_{\rm c}$} & \multicolumn{2}{l}{$\kern-1.5pt\Delta_>$\,(meV)} & \multicolumn{2}{l}{\!\!$2\Delta_>/k_{\rm B}T_{\rm c}$} & Experiment & Method or comment & Reference\\
$\phantom{\text{$x=12.5$\,\%}}$ & $\phantom{\text{(OD)}}$ & $\phantom{\text{Bridgman}}$ & $\phantom{\text{38}}$   & $\phantom{\text{0.25}}$ & $\phantom{\text{$\pm$\,0.03}}$ & $\phantom{\text{2.1}}$  & $\phantom{\text{$\pm$\,0.2}}$ & $\phantom{\text{10.67}}$ & $\phantom{\text{$\pm$\,0.04}}$ & $\phantom{\text{3.54}}$ & $\phantom{\text{$\pm$\,0.14}}$ & $\phantom{\text{magnetization}}$ & $\phantom{\text{$a\!b$-plane junction-average}}$ & $\phantom{\text{Kawano-Furukawa \textit{et~al.} [S19]}}$\vspace{-1.25em}\\

\bottomrule

\multicolumn{15}{c}{\!\bf $\kern-0.75pt$\hrulefill ~~122-family of ferropnictides\,$^{\strut}$ \hrulefill\!$\kern-0.75pt$}\\

\multicolumn{10}{l}{\!\bf Ba$_{1-x}$K$_x$Fe$_2$As$_2$, hole-doped (BKFA)}\\
\midrule
$x=25$\,\% & (UD) & FeAs-flux & 26   & 4.0 & $\pm$\,0.8 & 3.6  & $\pm$\,0.7 & ~~7.8 & $\pm$\,0.9 & 7.0  & $\pm$\,0.8 & ARPES & symmetrization & \citet{NakayamaSato11}\\
$x=40$\,\% & (OP) & ~\hfill»\hfill~ & 37   & 5.8 & $\pm$\,0.8 & 3.6  & $\pm$\,0.5 & 12.3 & $\pm$\,0.8 & 7.7  & $\pm$\,0.5 & ~~~~\,» & \qquad\quad» & »\hspace{4.5em} \\
$x=70$\,\% & (OD) & ~\hfill»\hfill~ & 22   & 4.4 & $\pm$\,0.8 & 4.6  & $\pm$\,0.9 & ~~7.9 & $\pm$\,0.8 & 8.3  & $\pm$\,0.9 & ~~~~\,» & \qquad\quad» & »\hspace{4.5em} \\
$x=29$\,\% & (UD) & Sn-flux   & 28   & 3.7 & $\pm$\,0.5$^\ast\!\!\!$ & 3.1  & $\pm$\,0.4$^\ast\!\!\!$ & \multicolumn{2}{c}{---~} & \multicolumn{2}{c}{---~~~} & PCAR              & $c$-axis Au junction & \citet{ZhangOh10}\\
$x=28$\,\% & ~\hfill»\hfill~ & \hfill»\hfill\quad~ & 31.5 & 2.3 &            & 1.7  &           & ~~9.8 &            & 7.2 &            & $^{75}$\!As-NMR   & spin-lattice relax. rate & \citet{MatanoLi09}\\
$x=32$\,\% & (OP) & FeAs-flux & 38.5 & 3.5 &            & 2.2  &            & 11  &            & 6.6 &            & calorimetry       & electronic specific heat  & \citet{PopovichBoris10}\\
$x=40$\,\% & ~\hfill»\hfill~ & ~\hfill»\hfill~ & 38 & 3.6 & $\pm$\,0.5 & 2.2 & $\pm$\,0.3 & ~~8.2 & $\pm$\,0.9 & 5.1 & $\pm$\,0.5 & STS & peak-to-peak distance & \citet{ShanWang11nphys}\\
$\phantom{x=}$» & ~\hfill»\hfill~ & ~\hfill»\hfill~   & 37   & 3.5 & $\pm$\,0.5 & 2.3  & $\pm$\,0.3 & \multicolumn{2}{c}{---~} & \multicolumn{2}{c}{---~~~} & PCAR            & single-gap BTK-fit & \citet{LuPark10}\\
$\phantom{x=}$» & ~\hfill»\hfill~ & ~\hfill»\hfill~ & ~» & 3.3 & $\pm$\,1.1$^\ast\!\!\!$ & 2.1  & $\pm$\,0.7$^\ast\!\!\!$ & ~~7.6  & $\pm$\,0.9$^\ast\!\!\!$ & 4.8 & $\pm$\,0.6$^\ast\!\!\!$ & STS & two-band model & \citet{ShanWang11}\\
$\phantom{x=}$» & ~\hfill»\hfill~ & ~\hfill»\hfill~ & ~»   & 6.4 & $\pm$\,0.5 & 4.0  & $\pm$\,0.3 & 11.3 & $\pm$\,1.0 & 7.0 & $\pm$\,0.6 & ARPES ($k_z$-resolved),~\hspace{-5em}      & \hspace{3em}symmetrization & \citet{XuHuang11}\\
$\phantom{x=}$» & ~\hfill»\hfill~ & ~\hfill»\hfill~ & ~»   & 6.0 & $\pm$\,1.0 & 3.7  & $\pm$\,0.6 & 12.0& $\pm$\,1.0 & 7.5 & $\pm$\,0.6 & ARPES             & symmetrization       & \citet{DingRichard08}\\
$\phantom{x=}$» & ~\hfill»\hfill~ & ~\hfill»\hfill~ & ~» & 6.0 & $\pm$\,1.5 & 3.7  & $\pm$\,1.0 & 13  & $\pm$\,2   & 8.1 & $\pm$\,1.2 & ARPES\,+\,STS     & \qquad\quad»       & \citet{WrayQian08}\\
$\phantom{x=}$» & ~\hfill»\hfill~ & ~\hfill»\hfill~ & ~» & 5.8 & $\pm$\,0.8 & 3.6  & $\pm$\,0.5 & 12.0& $\pm$\,0.8 & 7.5 & $\pm$\,0.4 & ARPES             & \qquad\quad»       & \citet{NakayamaSato09}\\
$\phantom{x=}$» & ~\hfill»\hfill~ & ~\hfill»\hfill~ & ~» & \multicolumn{2}{c}{---~~~} & \multicolumn{2}{c}{---~~~~} & 12.5& $\pm$\,2.0 & 7.8 & $\pm$\,1.2 & optics            & reflectance          & \citet{LiHu08}\\
$\phantom{x=}$» & ~\hfill»\hfill~ & ~\hfill»\hfill~ & 36.2 & 2.0 & $\pm$\,0.3 & 1.3  & $\pm$\,0.2 & ~~8.9& $\pm$\,0.4 & 5.7 & $\pm$\,0.3 & magnetization    & lower critical field & \citet{RenWang08}\\
$\phantom{x=}$» & ~\hfill»\hfill~ & ~\hfill»\hfill~ & 35   & 7.5 & $\pm$\,1.5 & 5.0  & $\pm$\,1.0 & 11  & $\pm$\,1.5 & 7.3 & $\pm$\,1.0 & ARPES             & symmetrization       & \citet{ZhaoLiu08}\\
$\phantom{x=}$» & ~\hfill»\hfill~ & Sn-flux   & 32   & <\,4&            & <\,3\!\! &            & ~~9.2& $\pm$\,1.0 & 6.7 & $\pm$\,0.7 & \quad»        & Dynes-function fit   & \citet{EvtushinskyInosov09}\\
$\phantom{x=}$» & ~\hfill»\hfill~ & \hfill»\hfill\quad~ & ~» & 1.5 & $\pm$\,1.0 & 1.1  & $\pm$\,0.7 & ~~9.1& $\pm$\,1.0 & 6.6 & $\pm$\,0.7 & ARPES\,+\,$\mu$SR& penetration depth    & \citet{KhasanovEvtushinsky09}\\
$\phantom{x=}$» & ~\hfill»\hfill~ & Bridgman  & 36   & 3.5 &            & 2.3  &            & 12  &            & 7.7 &            & optics            & optical conductivity & \citet{KwonHong10}\\
$\phantom{x=}$» & ~\hfill»\hfill~ & polycryst.    & 38   & 6.8 & $\pm$\,0.3 & 4.1  & $\pm$\,0.2 & 12  &            & 7.3 &            & $\mu$SR           & penetration depth    & \citet{HiraishiKadono09}\\
$x=45$\,\% & (OD) & Sn-flux   & 27   & 2.7 & $\pm$\,0.7$^\ast\!\!\!$ & 2.3  & $\pm$\,0.6$^\ast\!\!\!$ & ~~9.2 & $\pm$\,0.5$^\ast\!\!\!$ & 7.9 & $\pm$\,0.4$^\ast\!\!\!$ & PCAR            & $a\!b$-plane junction-average & \citet{SzaboPribulova09}\\
$x=49$\,\% & ~\hfill»\hfill~ & \hfill»\hfill\quad~ & 25.5 & 3.1 & $\pm$\,0.7$^\ast\!\!\!$ & 2.8  & $\pm$\,0.6$^\ast\!\!\!$ & \multicolumn{2}{c}{---~} & \multicolumn{2}{c}{---~~~} & \quad» & $c$-axis Pb junction & \citet{ZhangOh10}\\
$x=55$\,\% & ~\hfill»\hfill~ & FeAs-flux & 32.7 & 3.3 &            & 2.3  &            & ~~6.8 &          & 4.8 &            & MSI               & penetration depth    & \citet{HashimotoShibauchi09}\\
$x=77$\,\% & ~\hfill»\hfill~ & Sn-flux   & 21   & 2.7 & $\pm$\,0.3$^\ast\!\!\!$ & 3.0  & $\pm$\,0.4$^\ast\!\!\!$ & \multicolumn{2}{c}{---~} & \multicolumn{2}{c}{---~~~} & PCAR              & $c$-axis Pb junction & \citet{ZhangOh10}\\
\bottomrule

\multicolumn{10}{l}{\!\bf KFe$_2$As$_2$$^{\strut}$, ~\,100\,\% hole-doped (K-122 or KFA)}\\
\midrule
\multicolumn{2}{c}{N/A} & FeAs-flux & ~~4.0 & \multicolumn{2}{c}{---~~~} & \multicolumn{2}{c}{---~~~~} & ~~0.93 & $\pm$\,0.12 & 5.4 & $\pm$\,0.7 & TDR & nodal-gap model & \citet{HashimotoSerafin10}\\
         &      & ~\hfill»\hfill~ & ~~3.6  & 0.23& $\pm$\,0.03 & 1.5 & $\pm$\,0.2 & ~~0.55 & $\pm$\,0.02 & 3.55 & $\pm$\,0.13 & SANS            & 3-gap model & \citet{KawanoFurukawa10}\\
         &      & polycryst.& ~~3.5  & 0.07&             & 0.46\!\!\! &           & ~~0.73 &             & 4.84 &              & $^{75}$\!As-NQR & fully gapped $s^\pm$ & \citet{FukazawaYamada09}\\
\bottomrule

\multicolumn{10}{l}{\!\bf RbFe$_2$As$_2$$^{\strut}$, \!100\,\% hole-doped (Rb-122)}\\
\midrule
\multicolumn{2}{c}{N/A} & polycryst. & ~~2.5 & 0.15 & $\pm$\,0.02 & 1.4  & $\pm$\,0.2  & ~~0.49 & $\pm$\,0.04 & 4.5 & $\pm$\,0.4 & $\mu$SR         & penetration depth & \citet{ShermadiniKanter10}\\
\bottomrule

\multicolumn{10}{l}{\!\bf Ba(Fe$_{1-x}$Co$_x$)$_2$As$_2$$^{\strut}$, electron-doped (BFCA)}\\
\midrule
$x=7.0$\,\%  & (OP) & FeAs-flux & 22   & \multicolumn{2}{c}{---~~~} & \multicolumn{2}{c}{---~~~~} & ~~7.0 & $\pm$\,2.4$^\ast\!\!\!$ & 7.4 & $\pm$\,2.5$^\ast\!\!\!$ & STS   & peak-to-peak distance& \citet{MasseeHuang09}\\
$\phantom{x=}$» & ~\hfill»\hfill~ & ~\hfill»\hfill~ & 23   & \multicolumn{2}{c}{---~~~} & \multicolumn{2}{c}{---~~~~} & ~~5.5 & $\pm$\,0.5$^\ast\!\!\!$ & 5.5 & $\pm$\,0.5$^\ast\!\!\!$ & PCAR         & $c$-axis Pt junction & \citet{SamuelyPribulova09}\\
$\phantom{x=}$» & ~\hfill»\hfill~ & ~\hfill»\hfill~ & 24.5 & \multicolumn{2}{c}{---~~~} & \multicolumn{2}{c}{---~~~~} & ~~7.3 &                         & 6.9 &                         & STS          & peak-to-peak distance & \citet{NishizakiNakajima11}\\
$x=7.5$\,\%  & ~\hfill»\hfill~ & ~\hfill»\hfill~ & 25.5 & 4.5  & $\pm$\,1.0 & 4.1  & $\pm$\,0.9 & ~~6.7 & $\pm$\,1.0 & 6.1 & $\pm$\,0.9 & ARPES        & symmetrization       & \citet{TerashimaSekiba09}\\
$\phantom{x=}$» & ~\hfill»\hfill~ & ~\hfill»\hfill~ & 25   & 3.1  & $\pm$\,0.2 & 2.9  & $\pm$\,0.2 & ~~7.4 & $\pm$\,0.3 & 6.9 & $\pm$\,0.3 & optics & optical conductivity & \citet{TuLi10}\\
$x=10$\,\%   & ~\hfill»\hfill~ & ~\hfill»\hfill~ & 24.5 & 4.4  & $\pm$\,0.6 & 4.2  & $\pm$\,0.6 & ~~9.9 & $\pm$\,1.2     & 9.4&$\pm$\,1.1 & PCAR  & $a\!b$-plane, BTK-fit & \citet{TortelloDaghero10}\\
$\phantom{x=}$» & ~\hfill»\hfill~ & ~\hfill»\hfill~ & 25.3 & \multicolumn{2}{c}{---~~~} & \multicolumn{2}{c}{---~~~~} & ~~6.3 & $\pm$\,1.7$^\ast\!\!\!$ & 5.8&$\pm$\,1.6$^\ast\!\!\!$ & STS  & peak-to-peak distance    & \citet{YinZech09}\\
$x=7.4$\,\%  & (OD) & ~\hfill»\hfill~ & 22.5 & 1.5  &            & 1.6 &       & ~~3.7 &          & 3.8 &       & $\mu$SR     & penetration depth    & \citet{WilliamsAczel09}\\
$x=7.5$\,\%  & ~\hfill»\hfill~ & ~\hfill»\hfill~ & 21.4 & 1.75 &            & 1.9 &             & ~~4.1 &          & 4.4 &            & calorimetry & electronic specific heat & \citet{HardyWolf10}\\
$x=4.0$\,\%  & (UD) & ~\hfill»\hfill~ & ~~5.8& 0.38 &            & 1.5 &             & ~~0.86\!\!\!&    & 3.4 &            & \qquad\,» & \qquad\qquad» & \citet{HardyBurger10}\\
$x=4.5$\,\%  & ~\hfill»\hfill~ & ~\hfill»\hfill~ & 13.3 & 0.89 &            & 1.5 &             & ~~2.2 &          & 3.8 &            & \qquad\,» & \qquad\qquad» & »\hspace{4.5em} \\
$x=5.0$\,\%  & ~\hfill»\hfill~ & ~\hfill»\hfill~ & 19.5 & 1.36 &            & 1.6 &             & ~~3.5 &          & 4.2 &            & \qquad\,» & \qquad\qquad» & »\hspace{4.5em} \\
$x=5.5$\,\%  & ~\hfill»\hfill~ & ~\hfill»\hfill~ & 21.5 & 1.84 &            & 2.0 &             & ~~4.4 &          & 4.7 &            & \qquad\,» & \qquad\qquad» & »\hspace{4.5em} \\
$x=5.7$\,\%  & (OP) & ~\hfill»\hfill~ & 24.4 & 1.94 &            & 1.9 &             & ~~5.2 &          & 5.0 &            & \qquad\,» & \qquad\qquad» & »\hspace{4.5em} \\
$x=6.0$\,\%  & ~\hfill»\hfill~ & ~\hfill»\hfill~ & 24.2 & 1.94 &            & 1.8 &             & ~~5.0 &          & 4.8 &            & \qquad\,» & \qquad\qquad» & »\hspace{4.5em} \\
$x=6.5$\,\%  & (OD) & ~\hfill»\hfill~ & 23.8 & 1.78 &            & 1.7 &             & ~~4.6 &          & 4.5 &            & \qquad\,» & \qquad\qquad» & »\hspace{4.5em} \\
$x=7.5$\,\%  & ~\hfill»\hfill~ & ~\hfill»\hfill~ & 22.9 & 1.81 &            & 1.8 &             & ~~4.4 &          & 4.5 &            & \qquad\,» & \qquad\qquad» & »\hspace{4.5em} \\
$x=7.6$\,\%  & ~\hfill»\hfill~ & ~\hfill»\hfill~ & 21.5 & 1.84 &            & 2.0 &             & ~~3.9 &          & 4.2 &            & \qquad\,» & \qquad\qquad» & »\hspace{4.5em} \\
$x=9.0$\,\%  & ~\hfill»\hfill~ & ~\hfill»\hfill~ & 20.7 & 1.62 &            & 1.8 &             & ~~3.8 &          & 4.3 &            & \qquad\,» & \qquad\qquad» & »\hspace{4.5em} \\
$x=11.0$\,\% & ~\hfill»\hfill~ & ~\hfill»\hfill~ & 13.0 & 0.89 &            & 1.6 &             & ~~2.0 &          & 3.6 &            & \qquad\,» & \qquad\qquad» & »\hspace{4.5em} \\
$x=11.3$\,\% & ~\hfill»\hfill~ & ~\hfill»\hfill~ & 11.0 & 0.83 &            & 1.7 &             & ~~1.75\!\!\!&    & 3.7 &            & \qquad\,» & \qquad\qquad» & »\hspace{4.5em} \\
$x=11.6$\,\% & ~\hfill»\hfill~ & ~\hfill»\hfill~ & ~~9.4& 0.54 &            & 1.3 &             & ~~1.27\!\!\!&    & 3.1 &            & \qquad\,» & \qquad\qquad» & »\hspace{4.5em} \\
$x=12.0$\,\% & ~\hfill»\hfill~ & ~\hfill»\hfill~ & ~~5.1& 0.25 &            & 1.1 &             & ~~0.67\!\!\!&    & 3.1 &            & \qquad\,» & \qquad\qquad» & »\hspace{4.5em} \\
$x=6.0$\,\%  & (UD) & ~\hfill»\hfill~ & 14 & 4 & $\pm$\,2$^\ast\!\!\!$ & 7 & $\pm$\,3$^\ast\!\!\!$ & ~~8 & $\pm$\,2$^\ast\!\!\!$ & \!\!\!13 & $\pm$\,3$^\ast\!\!\!$ & STS   & Dynes-function fit & \citet{TeagueDrayna11}\\
$x=12.0$\,\% & (OD) & ~\hfill»\hfill~ & 20 & 5 & $\pm$\,2$^\ast\!\!\!$ & 6 & $\pm$\,3$^\ast\!\!\!$ & 10 & $\pm$\,2$^\ast\!\!\!$ & \!\!\!11 & $\pm$\,3$^\ast\!\!\!$ & ~~\,» & \qquad\qquad» & »\hspace{4.5em} \\
$x=10$\,\%   & ~\hfill»\hfill~ & thin film & ~» & 1.85  & $\pm$\,0.15 & 2.1  & $\pm$\,0.2 & \!\!$\geq$\,3.5 &                 & $\kern-1em\geq$\,4.0 &                         & optics       & optical conductivity & \citet{MaksimovKarakozov11}\\
$x=6.5$\,\%  & (OP) & FeAs-flux & 24.5 & 3.3  &            & 3.1  &            & ~~5.0 &                         & 4.7 &                         & \quad» & \qquad\qquad» & \citet{KimRossle10}\\
$x=4.9$\,\%  & (UD) & ~\hfill»\hfill~ & 15.8 & 0.8 & & 1.2 & & ~~3.0 & & 4.4 & & MFM & penetration depth & \citet{LuanLippman11}\\
$x=5.1$\,\%  & ~\hfill»\hfill~ & ~\hfill»\hfill~ & 18.6 & 1.1 & & 1.4 & & ~~3.7 & & 4.6 & & ~~\,» & \qquad\qquad» & »\hspace{4.5em} \\
$x=7.0$\,\%  & (OP) & ~\hfill»\hfill~ & 22.4 & 2.5 & & 2.6 & & ~~6.4 & & 6.6 & & ~~\,» & \qquad\qquad» & »\hspace{4.5em} \\
$x=8.5$\,\%  & (OD) & ~\hfill»\hfill~ & 19.6 & 1.0 & & 1.2 & & ~~3.2 & & 3.8 & & ~~\,» & \qquad\qquad» & »\hspace{4.5em} \\
$x=11$\,\%  & ~\hfill»\hfill~ & ~\hfill»\hfill~ & 13.5 & 0.7 & & 1.2 & & ~~2.0 & & 3.4 & & ~~\,» & \qquad\qquad» & »\hspace{4.5em} \\

\bottomrule

\multicolumn{15}{r}{\sl Continued on next page\!\!$^{\strut}$\vspace{-4em}}
\end{tabular}}\enlargethispage{2pt}\clearpage

{\flushleft\footnotesize
\begin{tabular}[c]{@{}l@{~}l@{~~}l@{~~}l@{~~}l@{\hspace{1pt}}l@{~~}l@{\hspace{1pt}}l@{\quad}l@{\hspace{1pt}}l@{~~}l@{\hspace{1pt}}l@{~~}l@{~~}l@{\hspace{-3.2em}}r@{}}
\multicolumn{15}{r}{\sl Continued from previous page\!\!}\\
\toprule
\multicolumn{2}{l}{\!Doping level} & sample & $\hspace{-0.5pt}T_{\rm c}$\,(K)\!\! & \multicolumn{2}{l}{$\Delta_<$\,(meV)} & \multicolumn{2}{l}{\!\!$2\Delta_</k_{\rm B}T_{\rm c}$} & \multicolumn{2}{l}{$\kern-1.5pt\Delta_>$\,(meV)} & \multicolumn{2}{l}{\!\!$2\Delta_>/k_{\rm B}T_{\rm c}$} & Experiment & Method or comment & Reference\\
$\phantom{\text{$x=12.5$\,\%}}$ & $\phantom{\text{(OD)}}$ & $\phantom{\text{Bridgman}}$ & $\phantom{\text{38}}$   & $\phantom{\text{0.25}}$ & $\phantom{\text{$\pm$\,0.03}}$ & $\phantom{\text{2.1}}$  & $\phantom{\text{$\pm$\,0.2}}$ & $\phantom{\text{10.67}}$ & $\phantom{\text{$\pm$\,0.04}}$ & $\phantom{\text{3.54}}$ & $\phantom{\text{$\pm$\,0.14}}$ & $\phantom{\text{magnetization}}$ & $\phantom{\text{$a\!b$-plane junction-average}}$ & $\phantom{\text{Kawano-Furukawa \textit{et~al.} [S19]}}$\vspace{-1.25em}\\

\bottomrule

\multicolumn{10}{l}{\!\bf EuFe$_2$(As$_{1-x}$P$_x$)$_2$$^{\strut}$, isovalently substituted (EFAP)}\\
\midrule
$x=18$\,\% & (OP) & Bridgman & 28 & \multicolumn{2}{c}{---~~~} & \multicolumn{2}{c}{---~~~~} & ~~4.7 &  & 3.8 &  & optics & optical conductivity & \citet{WuChanda11}\\
\bottomrule

\multicolumn{10}{l}{\!\bf Sr(Fe$_{1-x}$Co$_x$)$_2$As$_2$$^{\strut}$, electron-doped (SFCA)}\\
\midrule
$x=7.5$\,\% & (OP)  & Sn-flux   & 19.5 & 1.4 &            & 1.7  &             & ~~8.6 &            & 10.2 &            & STS         & peak-to-peak distance & \citet{ParkKhim11}\\
$x=12.5$\,\% & (OD) & FeAs-flux & 13.3 & 1.3 & $\pm$\,0.3 & 2.3  & $\pm$\,0.5  & ~~3.7 & $\pm$\,0.4 & 6.5 & $\pm$\,0.7 & $\mu$SR         & penetration depth & \citet{KhasanovMaisuradze09}\\
$x=13$\,\%   & ~\hfill»\hfill~ & ~\hfill»\hfill~ & 15.5 & 1.8 & $\pm$\,0.3$^\ast\!\!\!$ & 2.7  & $\pm$\,0.5$^\ast\!\!\!$  &  \multicolumn{2}{c}{---~} & \multicolumn{2}{c}{---~~~} & PCAR & $c$-axis Pb\,\&\,Au junctions & \citet{ZhangOh10}\\
\bottomrule

\multicolumn{10}{l}{\!\bf BaNi$_2$As$_2$$^{\strut}$, \!100\,\% electron-doped (BNA)}\\
\midrule
\multicolumn{2}{c}{N/A} & Pb-flux & ~~0.68\!\!\!   & \multicolumn{2}{c}{---~~~} & \multicolumn{2}{c}{---~~~~} & ~~0.095\hspace{-2em} &  & 3.24 &  & calorimetry         & electronic specific heat & \citet{KuritaRonning09}\\
\bottomrule
\\
\multicolumn{15}{c}{\!\bf $\kern-0.75pt$\hrulefill ~~1111-family of ferropnictides~\,\hrulefill\!$\kern-0.75pt$}\\

\multicolumn{2}{l}{\!\bf LaFeAsO$_{1-x}$F$_{x}$,} & \multicolumn{10}{l}{\!\!\bf electron-doped (La-1111)}\\
\midrule
$x=8$\,\%    & (UD) & polycryst. & 23 & 3.0 &            & 3.0  &             & ~~7.5 &           & 7.5   &            & $^{75}$\!As-NQR   & spin-lattice relax. rate & \citet{KawasakiShimada08}\\
$x=10$\,\%   & (OP) & ~\hfill»\hfill~ & 26 & 3.9 & $\pm$\,0.7 & 3.5  & $\pm$\,0.6  & \multicolumn{2}{c}{---~} & \multicolumn{2}{c}{---~~~} & PCAR            & BTK-fit & \citet{ShanWang08}\\
$\phantom{x=}$» & ~\hfill»\hfill~ & ~\hfill»\hfill~ & ~» & 3.4 & $\pm$\,0.5 & 3.0  & $\pm$\,0.5  & \multicolumn{2}{c}{---~} & \multicolumn{2}{c}{---~~~} & calorimetry     & electronic specific heat & \citet{MuZhu08}\\
$\phantom{x=}$» & ~\hfill»\hfill~ & ~\hfill»\hfill~ & ~» & 4.0 & $\pm$\,0.6 & 3.6  & $\pm$\,0.5  & \multicolumn{2}{c}{---~} & \multicolumn{2}{c}{---~~~} & magnetization     & lower critical field, $d$-wave fit\hspace{-5em} & \citet{RenWang08a}\\
$\phantom{x=}$» & ~\hfill»\hfill~ & ~\hfill»\hfill~ & 27 & 3.8 & $\pm$\,0.4 & 3.3  & $\pm$\,0.3  & 10.0 & $\pm$\,0.6 & 8.5 & $\pm$\,0.5 & PCAR            & generalized BTK-fit & \citet{GonnelliDaghero09}\\
\bottomrule

\multicolumn{2}{l}{\!\bf PrFeAsO$_{1-x}$F$_{x}$$^{\strut}$,} & \multicolumn{10}{l}{\!\!\bf electron-doped (Pr-1111)}\\
\midrule
$x=11$\,\%   & (UD) & polycryst. & 45 & 4.3 &            & 2.2  &             & 13.7 &           & 7.1   &            & \multicolumn{2}{l}{$^{75}$\!As- \& \!$^{19}$F\/-NMR} & \citet{MatanoRen08}\\
\bottomrule

\multicolumn{2}{l}{\!\bf NdFeAsO$_{1-x}$F$_{x}$$^{\strut}$,} & \multicolumn{10}{l}{\!\!\bf electron-doped (Nd-1111)}\\
\midrule
$x=10$\,\%   & (OP) & polycryst. & 51 & 5.1 & $\pm$\,0.2$^\ast\!\!\!$ & 2.6  & $\pm$\,0.1$^\ast\!\!\!$      & 11.7 & $\pm$\,1.2$^\ast\!\!\!$ & 5.7   & $\pm$\,0.5$^\ast\!\!\!$    & PCAR & Pt junctions & \citet{SamuelySzabo09}\\
$\phantom{x=}$»   & ~\hfill»\hfill~ & sol.-state & 53 & \multicolumn{2}{c}{---~~~} & \multicolumn{2}{c}{---~~~~} & 15 & $\pm$\,1.5 & 6.6   & $\pm$\,0.7    & ARPES & symmetrization & \citet{KondoSantanderSyro08}\\

\bottomrule

\multicolumn{2}{l}{\!\bf SmFeAsO$_{1-x}$F$_{x}$$^{\strut}$,} & \multicolumn{10}{l}{\!\!\bf electron-doped (Sm-1111)}\\
\midrule
$x=20$\,\%  & (OP) & monocryst.\! & 51.2\!\!\! & 6.45 & $\pm$\,0.25 & 3.0  & $\pm$\,0.2  & 16.6 & $\pm$\,1.6     & 7.7 & $\pm$\,0.9  & PCAR & Au contact, BTK fit & \citet{KarpinskiZhigadlo09}\\
$\phantom{x=}$»  & ~\hfill»\hfill~ & ~\hfill»\hfill~ & 49.5\!\!\! & 8.0 &  & 3.7  &  & \multicolumn{2}{c}{---~} & \multicolumn{2}{c}{---~~~} & TRS & photoinduced reflectivity & \citet{MerteljKabanov09}\\
$x=10$\,\%  & ~\hfill»\hfill~ & polycryst. & 51.5\!\!\! & 3.7 & $\pm$\,0.4 & 1.7  & $\pm$\,0.2    & 10.5 & $\pm$\,0.5     & 4.7 & $\pm$\,0.2  & PCAR & Pt/Ir or Au junctions & \citet{WangShan09}\\
$x=20$\,\%  & ~\hfill»\hfill~ & ~\hfill»\hfill~ & 52 & 6.15 & $\pm$\,0.45 & 2.7  & $\pm$\,0.2    & 18 & $\pm$\,3 & 8.0   & $\pm$\,1.3    & \quad» & Ag-paste contact & \citet{DagheroTortello09}\\
$x=9$\,\%   & (UD) & ~\hfill»\hfill~ & 42 & 4.9 & $\pm$\,0.5 & 2.7  & $\pm$\,0.3      & 15 & $\pm$\,1 & 8.3   & $\pm$\,0.6 & \quad» & \qquad\quad» & »\hspace{4.5em} \\
$x=15$\,\%  & ~\hfill»\hfill~ & ~\hfill»\hfill~ & ~» & 6.7 & $\pm$\,0.15 & 3.7  & $\pm$\,0.1 & \multicolumn{2}{c}{---~} & \multicolumn{2}{c}{---~~~} & \quad» & Au junctions & \citet{ChenTesanovic08}\\
\bottomrule

\multicolumn{12}{l}{\!\bf SmFeAsO$_{1-x}$$^{\strut}$, ~\,oxygen-deficient (Sm-1111)}\\
\midrule
$x=15$\,\%  & (OP) & polycryst. & 52 & 8.25 & $\pm$\,0.25 & 3.7  & $\pm$\,0.1    & \multicolumn{2}{c}{---~} & \multicolumn{2}{c}{---~~~} & STS & $d$-wave model & \citet{MilloAsulin08}\\
\bottomrule

\multicolumn{2}{l}{\!\bf TbFeAsO$_{1-x}$F$_{x}$$^{\strut}$,} & \multicolumn{10}{l}{\!\!\bf electron-doped (Tb-1111)}\\
\midrule
$x=10$\,\%   & (UD) & polycryst. & 50 & 5.0 & $\pm$\,0.8  & 2.3  & $\pm$\,0.4    & ~~8.8 & $\pm$\,0.5  & 4.1   & $\pm$\,0.2 & PCAR              & Au junctions & \citet{YatesMorrison09}\\
\bottomrule
\\
\multicolumn{15}{c}{\!\bf $\kern-0.75pt$\hrulefill ~~111-family of ferropnictides~\,\hrulefill\!$\kern-0.75pt$}\\

\multicolumn{10}{l}{\!\bf Li$_{1+\delta}$FeAs, undoped (Li-111 or LFA)}\\
\midrule
\multicolumn{2}{c}{N/A} & self-flux & 18 & 1.0 & $\pm$\,0.5 & 1.3 & $\pm$\,0.6  & ~~3.2 &        & 4.1 &  & ARPES       & Dynes-function fit &        \citet{BorisenkoZabolotnyy10}\\
\multicolumn{2}{c}{}    & ~\hfill»\hfill~ & 17 & \multicolumn{2}{c}{---~~~} & \multicolumn{2}{c}{---~~~~} & ~~3.0 & $\pm$\,0.2 & 4.1 & $\pm$\,0.3 & SANS\hspace{0.5pt}+\hspace{0.5pt}ARPES\!\!\!\!\! & penetration depth &        \citet{InosovWhite10}\\
\multicolumn{2}{c}{}    & ~\hfill»\hfill~ & 16.9\!\!\! & 1.2 &  & 1.6 &    & ~~2.6 &        & 3.6 &  & calorimetry & electronic specific heat &        \citet{StockertAbdelHafiez10}\\
\multicolumn{2}{c}{}    & ~\hfill»\hfill~ & 16         & \multicolumn{2}{c}{---~~~} & \multicolumn{2}{c}{---~~~~} & \hspace{-3pt}$\sim$\,2.5 &        & 3.6 &  & STS & preliminary result & \citet{Hanaguri11}\\
\multicolumn{2}{c}{}    & ~\hfill»\hfill~ & 17 & 1.4 & $\pm$\,0.4 & 1.9 & $\pm$\,0.6  & ~~2.96 & $\pm$\,0.05 & 4.0 & $\pm$\,0.1 &  MSI        & penetration depth &        \citet{ImaiTakahashi10}\\
\multicolumn{2}{c}{}    & Bridgman & 17.5\!\!\! & 1.4 & $\pm$\,0.1 & 1.9 & $\pm$\,0.13\!\!\! & ~~2.9 & $\pm$\,0.2 & 3.8 & $\pm$\,0.3  & magnetization  & lower critical field, $\mathbf{H}\!\parallel c$ \hspace{-3em}& \citet{SongGhim11}\\
\multicolumn{2}{c}{}    & ~\hfill»\hfill~  & ~» & 1.2 & $\pm$\,0.1 & 1.6 & $\pm$\,0.13\!\!\! & ~~2.9 & $\pm$\,0.2 & 3.8 & $\pm$\,0.3  & \qquad~~~» & lower critical field, $\mathbf{H}\!\parallel a\!b$\hspace{-3em} & »\hspace{4.5em} \\
\multicolumn{2}{c}{}    & ~\hfill»\hfill~ & ~» & 1.7 & & 2.22\!\!\!\! & & ~~2.8 & & 3.77 & & TDR & penetration depth &        \citet{KimTanatar11}\\
\multicolumn{2}{c}{}    & polycryst.& 17 & 1.9 &            & 2.6 &             & ~~4.4 &            & 6.0 &            & $^{75}$\!As-NQR & spin-lattice relax. rate &        \citet{LiOoe10, LiKawasaki11}\\
\multicolumn{2}{c}{}    & ~\hfill»\hfill~ & 15 & 0.7 &            & 1.2 &             & ~~2.3 &        & 3.5 &  & calorimetry        & electronic specific heat &        \citet{WeiChen10}\\
\multicolumn{2}{c}{}    & grains & ~» & 0.6 & $\pm$\,0.13& 1.0 & $\pm$\,0.4  & ~~3.3 & $\pm$\,1.0 & 5.4 & $\pm$\,1.6 & magnetization        & lower critical field &        \citet{SasmalLv10}\\

\bottomrule

\multicolumn{10}{l}{\!\bf NaFe$_{1-x}$Co$_x$As$^{\strut}$, electron-doped (Na-111)}\\
\midrule
$x=5$\,\%  & (OD) & self-flux & 18 & \multicolumn{2}{c}{---~~~} & \multicolumn{2}{c}{---~~~~} & ~~6.5  & $\pm$\,0.5 & 8.3 & $\pm$\,0.6\!\!\! & ARPES & symmetrization & \citet{LiuRichard10}\\
\bottomrule

\multicolumn{15}{r}{\sl Continued on next page\!\!$^{\strut}$}
\end{tabular}}\clearpage

\begin{table*}[!]\flushleft\footnotesize
\begin{tabular}[c]{@{}l@{~}l@{~~}l@{~~}l@{~~}l@{\hspace{1pt}}l@{~~}l@{\hspace{1pt}}l@{\quad}l@{\hspace{1pt}}l@{~~}l@{\hspace{1pt}}l@{~~}l@{~~}l@{\hspace{-3.2em}}r@{}}
\multicolumn{15}{r}{\sl Continued from previous page\!\!}\\
\toprule
\multicolumn{2}{l}{\!Doping level} & sample & $\hspace{-0.5pt}T_{\rm c}$\,(K)\!\! & \multicolumn{2}{l}{$\Delta_<$\,(meV)} & \multicolumn{2}{l}{\!\!$2\Delta_</k_{\rm B}T_{\rm c}$} & \multicolumn{2}{l}{$\kern-1.5pt\Delta_>$\,(meV)} & \multicolumn{2}{l}{\!\!$2\Delta_>/k_{\rm B}T_{\rm c}$} & Experiment & Method or comment & Reference\\
$\phantom{\text{$x=12.5$\,\%}}$ & $\phantom{\text{(OD)}}$ & $\phantom{\text{monocryst.\!}}$ & $\phantom{\text{38}}$   & $\phantom{\text{0.25}}$ & $\phantom{\text{$\pm$\,0.03}}$ & $\phantom{\text{2.1}}$  & $\phantom{\text{$\pm$\,0.2}}$ & $\phantom{\text{10.67}}$ & $\phantom{\text{$\pm$\,0.04}}$ & $\phantom{\text{3.54}}$ & $\phantom{\text{$\pm$\,0.14}}$ & $\phantom{\text{magnetization}}$ & $\phantom{\text{$a\!b$-plane junction-average}}$ & $\phantom{\text{Kawano-Furukawa \textit{et~al.} [S19]}}$\vspace{-1.25em}\\

\bottomrule

\multicolumn{15}{c}{\!\bf $\kern-0.75pt$\hrulefill ~~arsenic-free Fe-based superconductors\,$^{\strut}$~\,\hrulefill\!$\kern-0.75pt$}\\

\multicolumn{10}{l}{\!\bf FeSe$_{1-x}$, chemically deficient}\\
\midrule
$x=15$\,\% & (OD) & polycryst.        & ~~8.3\!\!\!& 0.38 & $\pm$\,0.01 & 1.1  & $\pm$\,0.02\!\!\! & ~~1.60  & $\pm$\,0.02\!\!\!      & 4.45 & $\pm$\,0.06\!\!\! & $\mu$SR           & penetration depth    & \citet{KhasanovConder08}\\
\bottomrule

\multicolumn{10}{l}{\!\bf FeTe$_{1-x}$Se$_x$$^{\strut}$\!, isovalently substituted}\\
\midrule
$x=50$\,\% & (OP) & Bridgman   & 14.6& 0.51 & $\pm$\,0.03 & 0.8  & $\pm$\,0.05\!\!\! & ~~2.61  & $\pm$\,0.09\!\!\!      & 4.15 & $\pm$\,0.14\!\!\! & $\mu$SR           & penetration depth    & \citet{BendeleWeyeneth10}\\
$\phantom{x=}$» & ~\hfill»\hfill~ & polycryst. & 14.4& 0.87 & $\pm$\,0.06 & 1.4  & $\pm$\,0.1 & ~~2.6  & $\pm$\,0.1      & 4.2 & $\pm$\,0.2 & ~~\,»           & \qquad\qquad»    & \citet{BiswasBalakrishnan10}\\
$\phantom{x=}$» & ~\hfill»\hfill~ & self-flux  & 13.9& \multicolumn{2}{c}{---~~~} & \multicolumn{2}{c}{---~~~~} & ~~2.2 &         & 3.7 &            & calorimetry      & electronic specific heat    & \citet{GuntherDeisenhofer11}\\
$x=45$\,\% & ~\hfill»\hfill~ & unidirect. & 14  & 2.5  &             & 4.1  &                   & ~~5.1   &                        & 8.5  &                   & optics            & optical conductivity & \citet{HomesAkrap10}\\
$\phantom{x=}$» & ~\hfill»\hfill~ & solidificat.\!\!\! & 14.2& \multicolumn{2}{c}{---~~~} & \multicolumn{2}{c}{---~~~~} & ~~3.8   &                        & 6.2  &                   & PCAR              & $c$-axis Au junctions& \citet{ParkHunt10}\\
$x=43$\,\% & ~\hfill»\hfill~ & self-flux & 14.7  & 2.5  &             & 3.92\!\!\!\!\! &                   & ~~3.7   &                        & 5.84 &                   & calorimetry & electronic specific heat & \citet{HuLiu11}\\
$x=15$\,\% & ~\hfill»\hfill~ & \hfill»\hfill\quad~ & 14 & \multicolumn{2}{c}{---~~~} & \multicolumn{2}{c}{---~~~~} & ~~2.3  &  & 3.8 &   & STS           & Dynes-function fit & \citet{KatoMizuguchi09}\\
\bottomrule

\multicolumn{10}{l}{\!\bf Fe$_{1-x}$Mn$_x$Te$_{0.5}$Se$_{0.5}$$^{\strut}$}\\
\midrule
$x=2$\,\% & (OP) & self-flux  & 14.4& \multicolumn{2}{c}{---~~~} & \multicolumn{2}{c}{---~~~~} & ~~2.7 &         & 4.4 &            & calorimetry      & electronic specific heat    & \citet{GuntherDeisenhofer11}\\
\bottomrule

\multicolumn{15}{l}{\!\bf $A_x$(Fe$_{1-\delta}$Se)$_2$ ($A$\,=\,K,\,Rb,\,Cs)$^{\strut}$, heavily electron-doped (highest $T_{\rm c}$ among arsenic-free Fe-based superconductors)}\\
\midrule
K,~\, $x=0.7$\!\!\! & ~(UD)\!\!\! & Bridgman  & 28         & 1.5 &     & 1.3   &             & \multicolumn{2}{c}{---~} & \multicolumn{2}{c}{---~~~} & optics & optical conductivity & \citet{YuanDong11}\\
Tl$_{0.63}$K$_{0.37}$\!\!\! & ~~\hfill»\hfill~ & ~\hfill»\hfill~ & 29 & \multicolumn{2}{c}{---~~~} & \multicolumn{2}{c}{---~~~~} & ~~8.5 & $\pm$\,1.0 & 6.8 & $\pm$\,0.8 & ARPES & symmetrization & \citet{WangQian11}\\
Tl$_{0.45}$K$_{0.34}$\!\!\! & ~~\hfill»\hfill~ & ~\hfill»\hfill~ & 28 & \multicolumn{2}{c}{---~~~} & \multicolumn{2}{c}{---~~~~} & ~~8.0 & & 6.6 & & \quad~» & \qquad\quad» & \citet{ZhaoMou11}\\
K,~\, $x=0.7$\!\!\! & ~(OP)\!\!\! & ~\hfill»\hfill~  & 32 & \multicolumn{2}{c}{---~~~} & \multicolumn{2}{c}{---~~~~} & ~~9.0 & & 6.5 & & \quad~» & \qquad\quad» & »\hspace{4.5em} \\
Tl$_{0.58}$Rb$_{0.42}$\hspace{-9em} & ~~\hfill»\hfill~ & ~\hfill»\hfill~ & ~» & \multicolumn{2}{c}{---~~~} & \multicolumn{2}{c}{---~~~~} & 12.5 & $\pm$\,2.5 & 9.1 & $\pm$\,1.8 & \quad~» & \qquad\quad» & \citet{MouLiu11}\\
K,~\, $x=0.8$\!\!\! & ~~\hfill»\hfill~ & self-flux & 31.7\!\!\! & \multicolumn{2}{c}{---~~~} & \multicolumn{2}{c}{---~~~~} & 10.3 & $\pm$\,2 & 7.5 & $\pm$\,1.5 & \quad~» & \qquad\quad» & \citet{ZhangYang11}\\
\,»\qquad~» & ~~\hfill»\hfill~ & \hfill»\hfill\quad~ & 30 & \multicolumn{2}{c}{---~~~} & \multicolumn{2}{c}{---~~~~} & 10.8 & $\pm$\,2 & 7.6 & $\pm$\,1.0 & $^{77}$Se NMR & spin-lattice relax. rate & \citet{YuMa11}\\
\,»\qquad~» & ~~\hfill»\hfill~ & \hfill»\hfill\quad~ & ~» & \multicolumn{2}{c}{---~~~} & \multicolumn{2}{c}{---~~~~} & 10.3 & $\pm$\,2 & 7.9 & $\pm$\,1.5 & \qquad~» & \qquad\quad» & »\hspace{4.5em} \\
\bottomrule

\end{tabular}
\caption{Summary of the energy gap measurements in Fe-based superconductors. The gap values are obtained from the published results of point-contact Andreev-reflection (PCAR) or tunneling spectroscopy, scanning tunneling spectroscopy (STS), angle-resolved photoelectron spectroscopy (ARPES) and optical spectroscopy measurements that directly probe the electronic density of states, as well as indirectly from the calorimetric measurements of the electronic specific heat, magnetization measurements of the lower critical field ($H_{\rm c1}$), muon-spin-rotation ($\mu$SR), small-angle neutron scattering (SANS), microwave surface-impedance (MSI), tunnel-diode resonator (TDR), or magnetic force microscopy (MFM) measurements of the London penetration depth, from the nuclear-magnetic-resonance (NMR) or nuclear-quadrupolar-resonance (NQR) measurements of the spin-lattice relaxation rate, and from the time-resolved
femtosecond spectroscopy (FTS) via the temperature-dependence of the photoinduced reflectivity. The error values marked by an asterisk represent the spread of the gap values measured in different points on the sample or using different junctions. They can be therefore larger than the uncertainty of the average.\label{Tab:GapsFeAs}}
\end{table*}


\begin{table*}[!]\vspace{-2.5em}\flushleft\footnotesize
\begin{tabular}[c]{@{}l@{~}l@{}l@{~~}l@{~~}l@{\hspace{1pt}}l@{~~}l@{\hspace{1pt}}l@{\quad}l@{\hspace{1pt}}l@{~~}l@{\hspace{1pt}}l@{~~}l@{~~}l@{\hspace{-3.2em}}r@{}}
\toprule
\multicolumn{2}{l}{\!Compound} & sample & $T_{\rm c}$\,(K) & \multicolumn{2}{l}{$\kern-2pt\Delta_<$\,(meV)} & \multicolumn{2}{l}{\!\!$2\Delta_</k_{\rm B}T_{\rm c}$} & \multicolumn{2}{l}{$\kern-1.5pt\Delta_>$\,(meV)} & \multicolumn{2}{l}{\!\!$2\Delta_>/k_{\rm B}T_{\rm c}$} & Experiment & Method or comment & Reference\\
$\phantom{\text{$x=12.5$\,\%}}$ & $\phantom{\text{(OD)}}$ & $\phantom{\text{Bridgman}}$ & $\phantom{\text{38.5}}$   & $\phantom{\text{0.25}}$ & $\phantom{\text{$\pm$\,0.03}}$ & $\phantom{\text{2.1}}$  & $\phantom{\text{$\pm$\,0.2}}$ & $\phantom{\text{10.67}}$ & $\phantom{\text{$\pm$\,0.04}}$ & $\phantom{\text{3.54}}$ & $\phantom{\text{$\pm$\,0.14}}$ & $\phantom{\text{magnetization}}$ & $\phantom{\text{$a\!b$-plane junction-average}}$ & $\phantom{\text{Kawano-Furukawa \textit{et~al.} [S19]}}$\vspace{-1.25em}\\
\bottomrule

\multicolumn{15}{c}{\!\bf $\kern-0.75pt$\hrulefill ~~multiband superconductors\,$^{\strut}$ \hrulefill\!$\kern-0.75pt$}\\

\multicolumn{10}{l}{\!\bf MgB$_2$}\\
\midrule
                      & & polycryst. & 39.3 & 2.8 & $\pm$\,0.2 & 1.7  & $\pm$\,0.2   & ~~7  & $\pm$\,0.5      & 4.1 & $\pm$\,0.3 & PCAR          & Cu junction    & \citet{SzaboSamuely01}\\
                      & & monocryst. & 38.2 & 2.9 & $\pm$\,0.3 & 1.8 & $\pm$\,0.2   & ~~7.1& $\pm$\,0.5      & 4.3 & $\pm$\,0.3 & \quad»          & Ag paint or In junctions    & \citet{GonnelliDaghero02}\\
                      & & polycryst. & 38.7 & 3.5 & $\pm$\,0.4 & 2.1  & $\pm$\,0.3   & ~~7.5& $\pm$\,0.5      & 4.5 & $\pm$\,0.3 & STS           & Dynes-function fit    & \citet{GiubileoRoditchev02}\\
                      & & thin films & 40   & 2.3 &            & 1.3  &              & ~~7.1&                 & 4.1 &            & ~~\,»           & peak-to-peak distance & \citet{IavaroneKarapetrov05}\\
                      & & polycryst. & 38.8 & 2.7 &            & 1.6  &              & ~~6.2&                 & 3.7 &            & Raman         & 2-gap fit & \citet{ChenKonstantinovic01}\\
                      & & ~\hfill»\hfill~ & 36.5 & 1.7 & $\pm$\,0.2 & 1.1  & $\pm$\,0.2   & ~~5.6& $\pm$\,0.2      & 3.5 & $\pm$\,0.2 & PES           & Dynes-function fit & \citet{TsudaYokoya01}\\
                      & & monocryst. & 36   & 2.3 & $\pm$\,0.4 & 1.5  & $\pm$\,0.3   & ~~5.5& $\pm$\,0.4      & 3.5 & $\pm$\,0.3 & ARPES         & BCS-function fit & \citet{TsudaYokoya03}\\
                      & & ~\hfill»\hfill~ & 38   & 1.5 & $\pm$\,0.5 & 0.9  & $\pm$\,0.3   & ~~6.5& $\pm$\,0.5      & 3.9 & $\pm$\,0.3 & \quad~»         & \qquad\quad» & \citet{SoumaMachida03}\\
\multicolumn{3}{l}{\!neutron-irradiated} & 7\,--\,38\!\!\! &     &            & 2.0  & $\pm$\,0.3   &      &                 & 3.5 & $\pm$\,0.3 & \multicolumn{2}{l}{$\text{\!specific heat, transport and PCAR (review) \hspace{-30em}}$} & \citet{Xi08}\\
\bottomrule

\multicolumn{15}{l}{\!\bf Mg(B$_{1-x}$C$_x$)$_2$, Mg$_{1-x}$Al$_x$B$_2$ or Mg$_{1-x}$Mn$_x$B$_2$$^{\strut}$ (chemically substituted MgB$_2$)}\\
\midrule
\multicolumn{2}{l}{\!C-substituted}  & & & & & 1.5 & $\pm$\,0.5 & & & 4.0 & $\pm$\,0.3 & PCAR & review & \citet{GonnelliDaghero07}\\
\multicolumn{2}{l}{\!Al-substituted} & & & & & 2.1 & $\pm$\,0.5 & & & 4.2 & $\pm$\,0.3 & \quad» & \quad~» & »\hspace{4.5em} \\
\multicolumn{2}{l}{\!Mn-substituted} & & & & & 1.9 & $\pm$\,0.2 & & & 3.7 & $\pm$\,0.5 & \quad» & \quad~» & »\hspace{4.5em} \\
\bottomrule

\multicolumn{10}{l}{\!\bf 2$H$-NbSe$_2$$^{\strut}$}\\
\midrule
                      & & monocryst. & ~~7.2  & 0.2 & $\pm$\,0.2 & 0.7  & $\pm$\,0.7   & ~~1.2& $\pm$\,0.1      & 3.8 & $\pm$\,0.3 & ARPES         & BCS-function fit  & \citet{YokoyaKiss01}\\
                      & & ~\hfill»\hfill~ & \quad»  & \multicolumn{2}{c}{---~~~} & \multicolumn{2}{c}{---~~~~} & ~~0.8& $\pm$\,0.4      & 2.6 & $\pm$\,1.3 & \quad~»         & leading edge      & \citet{BorisenkoKordyuk09}\\
                      & & ~\hfill»\hfill~ & ~~7.1  & 0.4 & $\pm$\,0.1 & 1.2  & $\pm$\,0.2   & ~~1.0& $\pm$\,0.2      & 3.2 & $\pm$\,0.6 & TDR & penetration depth & \citet{FletcherCarrington07}\\
\bottomrule

\multicolumn{10}{l}{\!\bf YNi$_2$B$_2$C$^{\strut}$}\\
\midrule
& & monocryst. & 13.77  & 1.19&            & 2.0       &                           & ~~2.67    &                & 4.5 &          & calorimetry & electronic specific heat & \citet{HuangLin06}\\
& & ~\hfill»\hfill~ & 14.5  & 0.31& $\pm$\,0.06$^\ast\!\!\!$ & 1.6       & $\pm$\,0.3$^\ast\!\!\!$  & ~~2.0       & $\pm$\,0.2$^\ast\!\!\!$ & 3.2 & $\pm$\,0.3$^\ast\!\!\!$ & PCAR &   & \citet{MukhopadhyaySheet05}\\
& & ~\hfill»\hfill~ & 15.2  & 1.6 &                          & 2.4       &                          & ~~2.8       &                         & 4.3 &                         & INS & phonon line shapes & \citet{WeberKreyssig08}\\
\bottomrule

\multicolumn{10}{l}{\!\bf $R_2$Fe$_3$Si$_5$$^{\strut}$ ($R=\text{Lu, Sc}$) or Sc$_5$Ir$_4$Si$_{10}$}\\
\midrule
\multicolumn{2}{l}{\!Lu$_2$Fe$_3$Si$_5$} & monocryst. & ~~5.8  & 0.3 &            & 1.1  &              & ~~1.1&                 & 4.4 &            & calorimetry   & electronic specific heat & \citet{NakajimaNakagawa08}\\
\multicolumn{2}{l}{\!Sc$_2$Fe$_3$Si$_5$} & ~\hfill»\hfill~ & ~~4.8  & 0.35&            & 1.7  &              & ~~0.74&                & 3.53&            & \qquad~»   & \qquad\qquad~» & \citet{TamegaiNakajima08}\\
\multicolumn{2}{l}{\!Sc$_5$Ir$_4$Si$_{10}$} & ~\hfill»\hfill~ & ~~8.2 & 0.7 &          & 1.9  &              & ~~1.45&                & 4.1 &            & \qquad~»    & \qquad\qquad~» & »\hspace{4.5em} \\
\bottomrule

\multicolumn{10}{l}{\!\bf V$_3$Si$^{\strut}$}\\
\midrule
& & monocryst. & 16.5  & 1.36 &      & 1.9 &             & ~~2.6 &          & 3.6 &          & MSI & penetration depth  & \citet{NefyodovShuvaev05}\\
\bottomrule

\multicolumn{10}{l}{\!\bf Ba$_8$Si$_{46}$$^{\strut}$}\\
\midrule
& & polycryst. & ~~8.1  & 0.9 & $\pm$\,0.2 & 2.6 & $\pm$\,0.6        & ~~1.3 & $\pm$\,0.1        & 3.7 & $\pm$\,0.3 & tunneling & BCS-function fit & \citet{NoatCren10}\\
\bottomrule

\multicolumn{10}{l}{\!\bf Mo$_3$Sb$_7$$^{\strut}$}\\
\midrule
& & polycryst. & ~~2.2  & 0.24&            & 2.5       &              & ~~0.38   &                & 4.0 &          & calorimetry & electronic specific heat & \citet{TranMiller08}\\
& & ~\hfill»\hfill~ & \quad» & 0.26&            & 2.73\!\!\!\! &              & ~~0.43   &                & 4.54 &          & $\mu$SR & penetration depth & \citet{TranHillier08}\\
\bottomrule


\multicolumn{10}{l}{\!\bf PrOs$_4$Sb$_{12}$$^{\strut}$ (heavy-fermion superconductor)}\\
\midrule
& & monocryst. & ~~1.75  & 0.09&            & 1.15\!\!\!\! &              & ~~0.27&                & 3.5&            & calorimetry & thermal conductivity & \citet{SeyfarthBrison06}\\
\bottomrule
\\
\multicolumn{15}{c}{\!\bf $\kern-0.75pt$\hrulefill ~~single-band superconductors~\,\hrulefill\!$\kern-0.75pt$}\\

\multicolumn{10}{l}{\!\bf Nb (highest $T_{\rm c}$ among elemental superconductors)}\\
\midrule
& & polycryst. & ~~9.26 & \multicolumn{4}{c}{------} & ~~1.5 &               & 3.7&              & PES & Dynes-function fit & \citet{ChainaniYokoya00}\\
\bottomrule

\multicolumn{10}{l}{\!\bf Pb$^{\strut}$}\\
\midrule
& & monocryst. & ~~7.2 & \multicolumn{4}{c}{------} & ~~1.35 & $\pm$\,0.06   & 4.3& $\pm$\,0.2 & \multicolumn{2}{l}{$\text{\!neutron spin-echo~~~~ phonon lifetimes \hspace{-30em}}$} & \citet{AynajianKeller08}\\

\bottomrule

\multicolumn{10}{l}{\!\bf Ba$_{1-x}$K$_x$BiO$_3$$^{\strut}$}\\
\midrule
$x=40$\,\% & \hspace{-1ex}(OP) & thin films & 19    & \multicolumn{4}{c}{------} & ~~3.0& $\pm$\,0.2    & 3.7& $\pm$\,0.5 & PCAR & Au junction & \citet{SatoTakagi90}\\
$\phantom{x=}$» & \hspace{-1em}~\hfill»\hfill~ & monocryst. & 30.8  & \multicolumn{4}{c}{------} & ~~5.6& $\pm$\,0.7    & 4.2& $\pm$\,0.5 & optics & reflectivity & \citet{PuchkovTimusk94}\\
$\phantom{x=}$» & \hspace{-1em}~\hfill»\hfill~ & ~\hfill»\hfill~ & 30    & \multicolumn{4}{c}{------} & ~~6.0&               & 4.6&            & \quad\,» & infrared conductivity & \citet{MarsiglioCarbotte96}\\
\bottomrule

\multicolumn{10}{l}{\!\bf CeCoIn$_5$$^{\strut}$ (highest $T_{\rm c}$ among heavy-fermion superconductors)}\\
\midrule
& & monocryst. & ~~2.3  & \multicolumn{4}{c}{------} & ~~0.46&                & 4.6&            & PCAR & Au junctions & \citet{ParkGreene05}\\
\bottomrule

\multicolumn{10}{l}{\!\bf UPd$_2$Al$_3$$^{\strut}$ (heavy-fermion superconductor)}\\
\midrule
& & thin film & ~~1.8  & \multicolumn{4}{c}{------} & ~~0.24&                & 3.0&            & tunneling & Dynes-function fit & \citet{JourdanHuth99}\\
\bottomrule

\multicolumn{10}{l}{\!\bf Sr$_2$RuO$_4$$^{\strut}$ (presumably a spin-triplet superconductor)}\\
\midrule
& & monocryst. & ~~1.5 & \multicolumn{4}{c}{------} & ~~0.28&              & 4.3 &  & tunneling & BCS-function fit & \citet{SuderowCrespo09}\\
\bottomrule

\end{tabular}
\caption{Selected reports of the energy gap measurements in multiband superconductors known before the discovery of high-$T_{\rm c}$ superconductivity in ferropnictides, as well as in several single-band superconductors. For high-$T_{\rm c}$ cuprates, see the next table. \vspace{30em}\label{Tab:GapsOther}}
\end{table*}\clearpage


\begin{table*}[!]\vspace{-1pt}\footnotesize
\begin{tabular}[c]{@{}l@{~~}l@{~~}l@{~~~}l@{~~}l@{\hspace{1pt}}l@{~~}l@{\hspace{1pt}}l@{~~~~~}l@{~~~}l@{}r@{}}
\toprule
\multicolumn{2}{l}{\!Compound} & sample & $T_{\rm c}$\,(K) & \multicolumn{2}{l}{$\Delta$\,(meV)} & \multicolumn{2}{l}{\!\!$2\Delta/k_{\rm B}T_{\rm c}$} & Experiment & Method or comment & Reference\\
\bottomrule

\multicolumn{11}{c}{\!\bf $\kern-0.75pt$\hrulefill ~~hole-doped cuprates\,$^{\strut}$ \hrulefill\!$\kern-0.75pt$}\\

\multicolumn{11}{l}{\!\bf Bi$_2$Sr$_2$CaCu$_2$O$_{8+\delta}$ (Bi-2212 or BSCCO)}\\
\midrule
         & (OD) &            & 70 & 23   & $\pm$\,3   & ~~7.6 & $\pm$\,1.0 & STM + tunneling & review & \citet{YuLi09}\\
         & ~\hfill»\hfill~ &            & 87 & 33.5 & $\pm$\,3.5 & ~~8.9 & $\pm$\,1.0 & \qquad~~» & \quad~» & »\hspace{4.5em} \\
         & ~\hfill»\hfill~ &            & 86 & 33 & $\pm$\,4 & ~~8.8 & $\pm$\,1.0 & ARPES & symmetrization + fit & \citet{LeeVishik07}\\
         & (OP) &            & 92 & 36 & $\pm$\,2 & ~~9.0 & $\pm$\,0.5 & \quad~» & \qquad\qquad» & »\hspace{4.5em} \\
         & ~\hfill»\hfill~ & float.-zone & 91 & 32 & $\pm$\,3 & ~~8.1 & $\pm$\,0.8 & \quad~» & empirical fit & \citet{FedorovValla99}\\
Pb-BSCCO & (UD) & annealed   & 77 & 28 &          & ~~8.4 &            & \quad~» & leading edge & \citet{BorisenkoKordyuk02}\\
\bottomrule

\multicolumn{11}{l}{\!\bf Bi$_2$Sr$_2$Ca$_2$Cu$_3$O$_{10+\delta}$$^{\strut}$ (Bi-2223)}\\
\midrule
         & (OD) &            & \!\!\!110 & 43  & $\pm$\,5 & ~~9.1 & $\pm$\,1.0 & ARPES & backfolded dispersion & \citet{IdetaTakashima10}\\
\bottomrule

\multicolumn{11}{l}{\!\bf La$_{2-x}$Sr$_x$CuO$_4$$^{\strut}$ (LSCO)}\\
\midrule
         & (OD) & float.-zone & 26 & ~~8.0  & $\pm$\,1.0 & ~~7.1 & $\pm$\,0.9 &       & review & \citet{YuLi09}\\
         & ~\hfill»\hfill~ & \qquad» & 31 & 10.5  &          & ~~7.8 &             & optics & \quad~» & »\hspace{4.5em} \\
$x=0.15$ & (OP) & \qquad» & 39 & 17.5 & $\pm$\,1.5 & 10.3  & $\pm$\,0.9 & ARPES & leading edge & \citet{YoshidaHashimoto09}\\
\bottomrule

\multicolumn{11}{l}{\!\bf YBa$_2$Cu$_3$O$_{6+\delta}$$^{\strut}$ (YBCO)}\\
\midrule
$\delta=0.6$ & (UD) &        & 63 & 39.5 & $\pm$\,1.5 & 14.4  & $\pm$\,0.6 &       & review     & \citet{YuLi09}\\
$\delta=0.7$ & ~\hfill»\hfill~ &        & 67 & 39.5 & $\pm$\,1.5 & 13.6  & $\pm$\,0.5 &         & \quad~» & »\hspace{4.5em} \\
$\delta=0.85$ & ~\hfill»\hfill~ &       & 89 & 39.5 & $\pm$\,1.5 & 10.2  & $\pm$\,0.4 &         & \quad~» & »\hspace{4.5em} \\
\bottomrule

\multicolumn{11}{l}{\!\bf Y$_{1-x}$Ca$_x$Ba$_2$Cu$_3$O$_7$$^{\strut}$ (Ca-YBCO)}\\
\midrule
$x=0.10$ & (OD) &             & 85.5 & 33 & $\pm$\,4 & ~~8.9 & $\pm$\,1.1 &         & review     & \citet{YuLi09}\\
$x=0.15$ & ~\hfill»\hfill~ &             & 75   & 26 & $\pm$\,3 & ~~7.0 & $\pm$\,0.8 &         & \quad~» & »\hspace{4.5em} \\
$\phantom{x=}$» & ~\hfill»\hfill~ & flux method & 77   & 29 & $\pm$\,3 & ~~8.7 & $\pm$\,1.0 & ARPES & peak position & \citet{ZabolotnyyBorisenko07}\\
\bottomrule

\multicolumn{11}{l}{\!\bf HgBa$_2$Ca$_2$Cu$_3$O$_{8+\delta}$$^{\strut}$ (Hg-1223)}\\
\midrule
         & (OP) &            & 130 & 60  &  & 10.6 &   & optics & review & \citet{YuLi09}\\
\bottomrule

\multicolumn{11}{l}{\!\bf HgBa$_2$CuO$_{4+\delta}$$^{\strut}$ (Hg-1201)}\\
\midrule
         & (OP) &            & 96 & 44  & $\pm$\,4 & 10.6 & $\pm$\,1.0 &       & review & \citet{YuLi09}\\
         & (UD) &            & 90 & 22.4  &          & ~~5.8 & $\pm$\,1.0 & optics &        & \citet{YangHwang09}\\
\multicolumn{3}{l}{\hspace{2em}UD\,78\,K~--~OD\,42\,K} & \hspace{-1em}42\,--\,95\hspace{-1em} & &       & ~~6.4 & $\pm$\,1.0 & Raman &        & \citet{GuyardSacuto08}\\
\bottomrule

\multicolumn{11}{l}{\!\bf Tl$_2$Ba$_2$CuO$_{6+\delta}$$^{\strut}$ (Tl-2201)}\\
\midrule
         & (OP) &            & 92.5 & 43  & $\pm$\,4 & 10.7 & $\pm$\,1.0 &        & review & \citet{YuLi09}\\
         & (OD) &            & 90   & 37  &          & ~~9.5 &             & optics & \quad~» & »\hspace{4.5em} \\
         & ~\hfill»\hfill~ & & ~»   & 28  &          & ~~7.2 &             & \quad\,» & \quad~» & \hspace{-5em}\citet{SchachingerCarbotte00}\\
\bottomrule

\multicolumn{11}{c}{\!\bf $\kern-0.75pt$\hrulefill ~~electron-doped cuprates\,$^{\strut}$ \hrulefill\!$\kern-0.75pt$}\\

\multicolumn{11}{l}{\!\bf Nd$_{2-x}$Ce$_x$CuO$_{4-\delta}$ (NCCO)}\\
\midrule
$x=0.15$ & (OP) & float.-zone & 22 & ~~5.0 & $\pm$\,1.0 & ~~5.2 & $\pm$\,1.1 & ARPES & leading edge & \citet{SatoKamiyama01}\\
\bottomrule

\multicolumn{11}{l}{\!\bf Pr$_{2-x}$Ce$_x$CuO$_{4-\delta}$$^{\strut}$ (PCCO)}\\
\midrule
$x=0.13$ & (UD) &   & 17 & ~~2.5 & $\pm$\,0.4 & ~~3.5 & $\pm$\,0.5 & tunneling & Pb/I/PCCO junctions & \citet{DaganBeck07}\\
$x=0.15$ & (OP) &   & 19 & ~~3.3 & $\pm$\,0.3 & ~~4.0 & $\pm$\,0.4 & \quad~~\,» & \qquad\qquad» & »\hspace{4.5em} \\
$x=0.16$ & (OD) &   & 16 & ~~2.6 & $\pm$\,0.4 & ~~3.8 & $\pm$\,0.5 & \quad~~\,» & \qquad\qquad» & »\hspace{4.5em} \\
\bottomrule

\multicolumn{11}{l}{\!\bf Pr$_{1-x}$LaCe$_x$CuO$_{4-\delta}$$^{\strut}$ (PLCCO)}\\
\midrule
$x=0.11$ & (OP) & float.-zone & 26 & ~~2.5 & $\pm$\,0.2 & ~~2.2 & $\pm$\,0.2 & ARPES & leading edge & \citet{MatsuiTerashima05}\\
$x=0.12$ & ~\hfill»\hfill~ & \qquad» & 25 & ~~3.6 & $\pm$\,0.2 & ~~3.5 & $\pm$\,0.2 & tunneling & Pt/Ir junctions & \citet{GiubileoPiano10}\\
$\phantom{x=}$» & ~\hfill»\hfill~ & \qquad» & 24 & ~~7.2 & $\pm$\,1.2 & ~~6.9 & $\pm$\,1.2 & STS &                & \citet{NiestemskiKunwar07}\\
\multicolumn{2}{l}{\!\!$x=0.12$,\,0.15}& \qquad» & \hspace{-1em}13\,--\,24\hspace{-1em} &  &  & ~~3.6 & $\pm$\,0.2 & PCAR & BTK-fit & \citet{ShanHuang08}\\
\bottomrule

\multicolumn{11}{c}{\!\bf $\kern-0.75pt$\hrulefill ~~ruthenocuprates\,$^{\strut}$ \hrulefill\!$\kern-0.75pt$}\\

\multicolumn{11}{l}{\!\bf RuSr$_2$GdCu$_2$O$_8$ (Ru-1212)}\\
\midrule
\multicolumn{2}{l}{\!Ru-1212} & polycryst. & 30 & ~~2.8 & $\pm$\,0.2 & ~~2.2 & $\pm$\,0.2 & PCAR & Pt/Ir junctions & \citet{PianoBobba06}\\
$\phantom{x=}$» & & \qquad» & 27 & ~~6.0 & $\pm$\,0.5 & ~~5.1 & $\pm$\,0.4 & \quad» & \qquad\quad\,» & \citet{CalzolariDaghero06}\\
\bottomrule
\end{tabular}
\caption{Summary of the energy gap measurements in copper-oxide-based superconductors.\label{Tab:GapsCuprates}}
\end{table*}

\begin{table*}[h]\vspace{-1pt}\footnotesize
\begin{tabular}[c]{@{}l@{~~}l@{~~}l@{~~~}l@{~~}l@{\hspace{1pt}}l@{~~~}r@{}}
\toprule
\multicolumn{2}{l}{\!Compound} & sample & $T_{\rm c}$\,(K) & \multicolumn{2}{l}{${\scriptstyle\Delta}C/\gamma_{\rm n} T_{\rm c}$} & Reference\\
\bottomrule
\multicolumn{7}{l}{\!\bf Ba$_{1-x}$K$_x$Fe$_2$As$_2$$^{\strut}$, hole-doped (BKFA)}\\
\midrule
$x=32$\,\% & (OP) & FeAs-flux & 38.5 & 2.5 & & \citet{PopovichBoris10}\\
\bottomrule
\multicolumn{7}{l}{\!\bf KFe$_2$As$_2$$^{\strut}$, 100\,\% hole-doped (K-122 or KFA)}\\
\midrule
\multicolumn{2}{c}{N/A} & polycryst. & ~~3.5 & 0.6 & & \citet{FukazawaYamada09}\\
\bottomrule
\multicolumn{7}{l}{\!\bf Ba(Fe$_{1-x}$Co$_x$)$_2$As$_2$$^{\strut}$, electron-doped (BFCA)}\\
\midrule
$x=7.5$\,\%  & (OD) & FeAs-flux & 21.4 & 1.6 & & \citet{HardyWolf10}\\
$x=5.75$\,\% & (OP) & \quad~~~» & 24.3 & 1.6 & & \citet{HardyBurger10}\\
$x=5.5$\,\%  & (UD) & \quad~~~» & 22.9 & 1.5 & & »\hspace{4.5em} \\
\bottomrule
\multicolumn{7}{l}{\!\bf BaNi$_2$As$_2$$^{\strut}$, 100\,\% electron-doped (BNA)}\\
\midrule
\multicolumn{2}{c}{N/A} & Pb-flux & ~~0.68 & 1.31\!\!\! & & \citet{RonningKurita08}\\
\bottomrule
\multicolumn{7}{l}{\!\bf PrFePO$^{\strut}$}\\
\midrule
\multicolumn{2}{c}{N/A} & O$_2$-annealed & ~~3.6 & 1.4 & & \citet{BaumbachHamlin09}\\
\bottomrule
\multicolumn{7}{l}{\!\bf LaFePO$^{\strut}$}\\
\midrule
\multicolumn{2}{c}{N/A} & Sn-flux & ~~5.9 & 0.6 & $\pm$\,0.2 & \citet{AnalytisChu08}\\
\bottomrule
\multicolumn{7}{l}{\!\bf LaNiAsO$_{1-x}$F$_x$$^{\strut}$}\\
\midrule
$x=5.5$\,\%& (OP) & polycryst. & ~~3.8 & 1.9 &  & \citet{LiChen08}\\
\bottomrule
\multicolumn{7}{l}{\!\bf Li$_{1+\delta}$FeAs$^{\strut}$, undoped (Li-111 or LFA)}\\
\midrule
\multicolumn{2}{c}{N/A} & grains & 17 & 1.2 & $\pm$\,0.2 & \citet{WeiChen10}\\
\multicolumn{2}{c}{N/A} & self-flux & 16.9 & 1.24 &  & \citet{StockertAbdelHafiez10}\\
\bottomrule
\multicolumn{7}{l}{\!\bf FeTe$_{1-x}$Se$_x$$^{\strut}$}\\
\midrule
$x=43$\,\%& (OP) & self-flux & 14.7 & 2.11 & & \citet{HuLiu11}\\
\bottomrule
\multicolumn{7}{l}{\!\bf K$_x$(Fe$_{1-\delta}$Se)$_2$$^{\strut}$ (KFS)}\\
\midrule
$x=0.8$&  & Bridgman & 32 & 1.93 & & \citet{ZengShen11}\\
\bottomrule
\end{tabular}
\caption{Heat-capacity measurements of the specific-heat-jump ratio, ${\scriptstyle\Delta}C/\gamma_{\rm n} T_{\rm c}$, in iron arsenide superconductors (also see Fig.\,\ref{Fig:SpecificHeat}).\label{Tab:SpecificHeatJump}\vspace{-1em}}
\end{table*}\clearpage

\begin{table*}[h]\vspace{-1pt}\footnotesize
\begin{tabular}[c]{@{}l@{~~}l@{~~}l@{~~~}l@{~~}r@{\hspace{1.2pt}}l@{~~}r@{\hspace{1.2pt}}l@{~~~}r@{\hspace{1.2pt}}l@{~~}r@{\hspace{1.2pt}}l@{~~~}l@{\hspace{1.2pt}}l@{~~~}l@{\hspace{1.2pt}}l@{~~}l@{\hspace{1.2pt}}l@{~~~}r@{}}
\toprule
\multicolumn{2}{l}{\!Compound} & sample & $T_{\rm c}$\,(K) & \multicolumn{4}{c}{$\omega_{\rm res}$ (meV)} & \multicolumn{4}{c}{$\omega_{\rm res}/k_{\rm B}T_{\rm c}$} & \multicolumn{2}{c}{$2\Delta_>$\,(meV)} & \multicolumn{4}{c}{$\omega_{\rm res}/2\Delta_>$} & Reference\\
                             & &        &                  & \multicolumn{2}{c}{$\!q_z=0$} & \multicolumn{2}{c}{$\!q_z=\piup$} & \multicolumn{2}{c}{$\!q_z=0$} & \multicolumn{2}{c}{$\!q_z=\piup$} & & & \multicolumn{2}{c}{$\!\!q_z=0$} & \multicolumn{2}{c}{$\!\!q_z=\piup$} \\
\bottomrule

\multicolumn{16}{l}{\!\bf Ba$_{1-x}$K$_x$Fe$_2$As$_2$$^{\strut}$, hole-doped (BKFA)}\\
\midrule
$x=40$\,\% & (OP) & polycryst. & 38 & \multicolumn{2}{c}{---~} & 14.0 & $\pm$\,1.0 & \multicolumn{2}{c}{---~~\,} & 4.3&$\pm$\,0.3 & 22.9&$\pm$\,1.0 & \multicolumn{2}{c}{---~~\,} & 0.61&$\pm$\,0.05 & \citet{ChristiansonGoremychkin08}\\
$x=33$\,\% & ~\hfill»\hfill~ & self-flux  & ~» & 15.0 & $\pm$\,1.0 & 16.0 & $\pm$\,1.0 & 4.6&$\pm$\,0.3 & 4.9&$\pm$\,0.3 & 22.9&$\pm$\,1.0 & 0.66&$\pm$\,0.05 & 0.70&$\pm$\,0.05 & \citet{ZhangWang11}\\
\bottomrule

\multicolumn{16}{l}{\!\bf Ba(Fe$_{1-x}$Co$_x$)$_2$As$_2$$^{\strut}$, electron-doped (BFCA)}\\
\midrule
$x=4$\,\% & (UD) & self-flux  & 11 & \multicolumn{2}{c}{---~} & 4.5 & $\pm$\,0.5 & \multicolumn{2}{c}{---~~\,} & 4.7&$\pm$\,0.5 & ~~4.4&$\pm$\,0.9 & \multicolumn{2}{c}{---~~\,} & 1.0&$\pm$\,0.2 & \citet{ChristiansonLumsden09}\\
$x=4.7$\,\% & ~\hfill»\hfill~ & \quad~\,»  & 17 & 9.0 & $\pm$\,1.0 & 5.0 & $\pm$\,0.5 & 6.1 & $\pm$\,0.7 & 3.4&$\pm$\,0.4 & ~~4.7&$\pm$\,1.3 & 1.9&$\pm$\,0.6 & 1.1&$\pm$\,0.3 & \citet{PrattKreyssig10}\\
$x=7.5$\,\% & (OP) & \quad~\,»  & 25 & 9.7 & $\pm$\,0.5 & 9.0 & $\pm$\,0.5 & 4.5 & $\pm$\,0.3 & 4.2&$\pm$\,0.3 & 12.3&$\pm$\,0.9 & 0.79&$\pm$\,0.07 & 0.73&$\pm$\,0.07 & \citet{InosovPark10}\\
$x=8$\,\% & (OD) & \quad~\,»  & 22 & 8.6 & $\pm$\,0.5 & 8.6 & $\pm$\,0.5 & 4.5&$\pm$\,0.3 & 4.5&$\pm$\,0.3 & 10.8&$\pm$\,0.8 & 0.80&$\pm$\,0.07 & 0.70&$\pm$\,0.05 & \citet{LumsdenChristianson09}\\
\bottomrule

\multicolumn{16}{l}{\!\bf Ba(Fe$_{1-x}$Ni$_x$)$_2$As$_2$$^{\strut}$, electron-doped (BFNA)}\\
\midrule
$x=3.7$\,\% & (UD) & self-flux  & 12.2 & 7.0 & $\pm$\,0.8 & 5.0 & $\pm$\,0.5 & 6.7&$\pm$\,0.8 & 4.8&$\pm$\,0.5 & \multicolumn{6}{l}{unknown (no direct measurements)} & \citet{WangLuo10}\\
$x=4.5$\,\% & ~\hfill»\hfill~ & \quad~\,»  & 18 & 8.9 & $\pm$\,0.8 & 6.5 & $\pm$\,1.0 & 5.7&$\pm$\,0.5 & 4.2&$\pm$\,0.6 & \multicolumn{6}{l}{\hspace{7em}»} & \citet{ParkInosov10}\\
$x=5$\,\% & (OP) & \quad~\,»  & 20 & 9.1 & $\pm$\,0.4 & 7.2 & $\pm$\,0.5 & 5.3&$\pm$\,0.3 & 4.2&$\pm$\,0.3 & \multicolumn{6}{l}{\hspace{7em}»} & \citet{ChiSchneidewind09}\\
$\phantom{x=}$» & ~\hfill»\hfill~ & \quad~\,»  & ~» & 8.7 & $\pm$\,0.4 & 7.2 & $\pm$\,0.7 & 5.1&$\pm$\,0.3 & 4.2&$\pm$\,0.4 & \multicolumn{6}{l}{\hspace{7em}»} & \citet{LiChen09}\\
$\phantom{x=}$» & ~\hfill»\hfill~ & \quad~\,»  & ~» & 8.0 & $\pm$\,0.5 & \multicolumn{2}{c}{---~~} & 4.6&$\pm$\,0.3 & \multicolumn{2}{c}{---~~\,} & \multicolumn{6}{l}{\hspace{7em}»} & \citet{ZhaoRegnault10}\\
$x=7.5$\,\% & (OD) & \quad~\,»  & 15.5 & 8.0 & $\pm$\,2.0 & 6.0 & $\pm$\,0.5 & 6.0&$\pm$\,1.5 & 4.5&$\pm$\,0.4 & \multicolumn{6}{l}{\hspace{7em}»} & \citet{WangLuo10}\\
\bottomrule

\multicolumn{16}{l}{\!\bf BaFe$_2$(As$_{1-x}$P$_x$)$_2$$^{\strut}$, isovalently substituted (BFAP)}\\
\midrule
$x=35$\,\% & (OP) & polycryst. & 30 & \multicolumn{2}{c}{---~} & 11.5 & $\pm$\,1.5 & \multicolumn{2}{c}{---~~\,} & 4.5 & $\pm$\,0.6 & \multicolumn{6}{l}{unknown (no direct measurements)}& \citet{IshikadoNagai11}\\
\bottomrule

\multicolumn{16}{l}{\!\bf LaFeAsO$_{1-x}$F$_{x}$$^{\strut}$, electron-doped (La-1111)}\\
\midrule
$x=8$\,\% & (OP) & polycryst. & 29 & \multicolumn{2}{c}{---~} & 13.0 & $\pm$\,1.0 & \multicolumn{2}{c}{---~~\,} & 5.2 & $\pm$\,0.4 & 20.0 & $\pm$\,1.2 & \multicolumn{2}{c}{---~~~\,} & 0.65&$\pm$\,0.06 & \citet{ShamotoIshikado10}\\
\bottomrule

\multicolumn{16}{l}{\!\bf Li$_{1+\delta}$FeAs$^{\strut}$, undoped (Li-111 or LFA)}\\
\midrule
\multicolumn{2}{c}{N/A} & polycryst. & 17 & \multicolumn{2}{c}{---~} & 8.0 & $\pm$\,2.0 & \multicolumn{2}{c}{---~~\,} & 5.5 & $\pm$\,1.4 & ~~6.1 & $\pm$\,0.5 & \multicolumn{2}{c}{---~~~\,} & 1.3&$\pm$\,0.4 & \citet{TaylorPitcher11}\\
\bottomrule

\multicolumn{16}{l}{\!\bf FeTe$_{1-x}$Se$_{x}$$^{\strut}$, isovalently substituted (11-family)}\\
\midrule
$x=0.4$ & (OP) & self-flux & 14 & \multicolumn{2}{c}{---~} & 6.5 & $\pm$\,0.5 & \multicolumn{2}{c}{---~~\,} & 5.3 & $\pm$\,0.4 & ~~6.9 & $\pm$\,1.2 & \multicolumn{2}{c}{---~~\,} & 0.94&$\pm$\,0.18 &\citet{QiuBao09}\\
$\phantom{x=}$» & ~\hfill»\hfill~ & \quad~\,» & ~» & 6.0 & $\pm$\,0.5 & \multicolumn{2}{c}{---~~} & 5.0 & $\pm$\,0.4 & \multicolumn{2}{c}{---~~\,} & \multicolumn{2}{c}{»~~\,} & 0.87&$\pm$\,0.17 & \multicolumn{2}{c}{---~~~\,} & \citet{ArgyriouHiess10}\\
$x=0.5$ & ~\hfill»\hfill~ & unidirect. solidif.\!\! & ~» & 6.2 & $\pm$\,0.5 & \multicolumn{2}{c}{---~~} & 5.1 & $\pm$\,0.4 & \multicolumn{2}{c}{---~~\,} & \multicolumn{2}{c}{»~~\,} & 0.90&$\pm$\,0.17 & \multicolumn{2}{c}{---~~~\,} &\citet{WenXu10}\\
$\phantom{x=}$» & ~\hfill»\hfill~ & Bridgman & ~» & 6.5 & $\pm$\,0.5 & \multicolumn{2}{c}{---~~} & 5.3 & $\pm$\,0.4 & \multicolumn{2}{c}{---~~\,} & \multicolumn{2}{c}{»~~\,} & 0.94&$\pm$\,0.18 & \multicolumn{2}{c}{---~~~\,} &\citet{MookLumsden10, MookLumsden10a}\\
\bottomrule

\end{tabular}
\caption{Summary of the spin resonance energies ($\omega_{\rm res}$), corresponding onset energies of the particle-hole continuum ($2\Delta_>$), normalized resonance energies ($\omega_{\rm res}/k_{\rm B}T_{\rm c}$), and the $\omega_{\rm res}/2\Delta_>$ ratios in Fe-based superconductors.\label{Tab:Resonances}}
\end{table*}

\twocolumngrid

\bibliography{Multigap}\onecolumngrid\vfill

\vfill
\clearpage
\end{document}